\documentclass[runningheads]{llncs}
\usepackage[utf8]{inputenc}
\usepackage{tabularx}
\usepackage{longtable}
\usepackage{url}
\usepackage{graphicx}
\usepackage{comment}
\usepackage[dvipsnames]{xcolor}
\usepackage{amsmath}

\begin{document}

\title{IDPFilter: Mitigating Interdependent Privacy Issues in Third-Party Apps}
%\thanks{Supported by organization x.}}

\author{Shuaishuai Liu\inst{1} \and
Gergely Bicz{\'o}k\inst{1}\inst{2}}

\institute{CrySyS Lab, Dept. of Networked Systems and Services,\\ Budapest Univ. of Technology and Economics, Hungary\\
\email{\{sliu,biczok\}@crysys.hu}
\and
HUN-REN-BME Information Systems Research Group
}
\maketitle              % typeset the header of the contribution
\begin{abstract}

Third-party applications have become an essential part of today's online ecosystem, enhancing the functionality of popular platforms. However, the intensive data exchange underlying their proliferation has increased concerns about interdependent privacy (IDP). This paper provides a comprehensive investigation into the previously underinvestigated IDP issues of third-party apps. Specifically, first, we analyze the permission structure of multiple app platforms, identifying permissions that have the potential to cause interdependent privacy issues by enabling a user to share someone else's personal data with an app. Second, we collect datasets and characterize the extent to which existing apps request these permissions, revealing the relationship between characteristics such as the respective app platform, the app's type, and the number of interdependent privacy-related permissions it requests. Third, we analyze the various reasons IDP is neglected by both data protection regulations and app platforms and then devise principles that should be followed when designing a mitigation solution. Finally, based on these principles and satisfying clearly defined objectives, we propose IDPFilter, a platform-agnostic API that enables application providers to minimize collateral information collection by filtering out data collected from their users but implicating others as data subjects. We implement a proof-of-concept prototype, IDPTextFilter, that implements the filtering logic on textual data, and provide its initial performance evaluation with regard to privacy, accuracy, and efficiency.

\end{abstract}

\section{Introduction}
\label{sec:intro}

Third-party applications, commonly referred to as apps, occupy a significant place within the present-day Internet ecosystem. These apps enhance the functionality and features of popular platforms, including but not limited to mobile operating systems, social media networks, web browsers, and cloud services. The proliferation of third-party apps is predicated upon the sharing of data, which is a crucial aspect of their respective platforms. The vast, diverse, and constant data exchange, however, has given rise to increasingly pressing concerns regarding privacy.

Recently, various app platforms have undertaken the initiative to improve their privacy preservation mechanisms. For instance, since the release of iOS v14.5\footnote{\url{https://developer.apple.com/app-store/user-privacy-and-data-use/}}, applications have been required to obtain user consent before tracking user data from other applications and websites. Additionally, Android v12 implemented a privacy dashboard\footnote{ \url{https://www.androidauthority.com/android-privacy-dashboard-1233846/}} enabling users to monitor app permission usage with increased accuracy, while Android v13 introduced fine-grain data sharing, e.g., granting access to a single photo instead of an entire album or all photos. These developments are encouraging steps in the ongoing quest for enhanced transparency, user control, and overall privacy. However, they do not address the issue of interdependent privacy (IDP)~\cite{DBLP:conf/fc/BiczokC13}, where data shared with an app by a user may contain personal and potentially sensitive information about their friends, contacts, or colleagues; without their knowledge or control. This is particularly pertinent in the context of third-party apps.

In recent years, some of the most prominent cases of interdependent privacy have been associated with Facebook. The Cambridge Analytica scandal, which was widely reported in the media, serves as a notable example~\cite{DBLP:journals/compsec/SymeonidisBSPSP18}. In this instance, the app ``thisisyourdigitallife'' harvested 87 million Facebook profiles and used the data to create comprehensive personal psychological profiles. This information was then utilized to deliver personalized political advertisements to influence the outcome of the 2016 US presidential election (and others). The app exploited the mechanism of \emph{collateral information collection} (i.e., the information collection process enabled by privacy interdependence) on Facebook, where it was installed by $270,000$ users but was able to access the profiles of tens of millions of friends owing to the controversial design of permissions in Facebook's Graph API~\cite{DBLP:journals/compsec/SymeonidisBSPSP18}.

In February 2021, Facebook reached a settlement in a class action lawsuit involving the use of facial recognition technology in its photo tagging function. The lawsuit, which was initiated in Illinois in 2015, claimed that Facebook created and stored scans of users' faces without their permission. The photo tagging feature allowed users to tag their friends in photos uploaded to Facebook, thereby establishing a personal connection to the friends' profiles without their consent\footnote{\url{https://www.theguardian.com/technology/2021/feb/27/facebook-illinois-privacy-lawsuit-settlement}}. The lawsuit resulted in Facebook compensating over $1.6$ million users with a total settlement amount of \$$650$ million, which is one of the largest privacy-related settlements to date. This case serves as a clear illustration of interdependent privacy, where the uploading user consciously gave Facebook permission to display their photo, but the users who were tagged in the photo suffered a loss of privacy without even being aware of it.

In addition to high-profile incidents, there have also been a number of less publicized instances of interdependent privacy violations. In 2015, a security vulnerability allowed third-party applications to access Google+ user profile data, which was not discovered and patched until March 2018\footnote{\url{https://www.wsj.com/articles/google-exposed-user-data-feared} \\ \url{-repercussions-of-disclosing-to-public-1539017194}}. This bug enabled developers to also obtain non-public profile fields for both the user and their friends, affecting an estimated $500,000$ profiles. Unsurprisingly, Google chose not to disclose this issue to the public, not to diminish its momentary competitive advantage over Facebook. Another concerning example is the popular TrueCaller Android app, which is capable of blacklisting spam numbers. The app requires the uploading of the installing user's address book to its servers, which constitutes an interdependent privacy issue as noted by the Article 29 Working Party in 2017\footnote{\url{https://ec.europa.eu/newsroom/article29/items/610173}}. TrueCaller also allows users to tag unknown numbers after calls and upload them for other users to see, which, in 2019, jeopardized the cover of an investigative journalist in a non-friendly country. Fortunately, no harm came to the journalist and their sources; however, the situation posed a considerable threat to their physical safety\footnote{\url{https://privacyinternational.org/node/2997}}.

The above incidents naturally invite three research questions.
\begin{itemize}
\item \emph{RQ1}: \emph{are interdependent privacy issues pervasive among third-party app platforms?}
\item \emph{RQ2}: \emph{do actual apps request the permissions enabling collateral information collection on their respective platforms?}
\item \emph{RQ3}: \emph{is it possible to devise a technical solution to mitigate collateral information collection on third-party app platforms?}
\end{itemize}

In this paper, we answer all three research questions affirmatively. Specifically, the contribution of this paper is four-fold.

First, we conduct an analysis of the permission systems of multiple app platforms and identify those permissions that have the potential to cause interdependent privacy issues.
Second, we collect datasets from the respective app platforms and characterize the extent to which existing apps request these permissions. Our findings indicate that the category of the app is a reliable predictor for the number of interdependent privacy-related permissions that are requested.
Third, we discuss potential transparency and control measures for mitigating the issues above.
Finally and most importantly, we design IDPFilter, a platform-agnostic API that enables application providers to filter out data collected from their users but implicating other natural persons as data subjects. We also implement a proof-of-concept prototype, IDPTextFilter, and provide its initial evaluation.

The rest of this paper is organized as follows. Section \ref{sec:related} briefly surveys the related work. Section \ref{sec:permissions} analyzes the permission structure of multiple widely-used third-party app platforms (Android, browser extensions, Google Workspace, and Zoom Marketplace) and identifies permissions (potentially) invoking interdependent privacy. Section \ref{sec:data} introduces our dataset and characterizes the proliferation of such permissions requested by real existing apps from their respective users. Section \ref{sec:mitigation} discusses prospective transparency- and control-enhancing techniques and devises IDPFilter, a platform-agnostic filtering API able to partially mitigate interdependent privacy issues of third-party apps.  Section~\ref{sec:api} describes our proof-of-concept prototype IDPTextFilter. Finally, Section \ref{sec:conclusion} provides a discussion and concludes our paper.
\section{Related work}
\label{sec:related}
In this section, we provide a succinct overview of existing literature in the field of interdependent privacy and third-party applications.

\emph{Interdependent privacy} captures the networked characteristics of privacy-related decisions. Owing to this networked nature, the privacy of individuals is bound to be affected by the actions of others, e.g., Facebook users sharing the data of their friends~\cite{DBLP:conf/fc/BiczokC13}. In economic terms, unaware fellow users fall victim to a negative externality. Extending this interpretation, a data entry, seemingly concerning a single individual, may actually be also related to (multiple) others because of data correlation~\cite{DBLP:conf/ndss/OlteanuHDH18}. Note that the same concept is known under different monikers, such as collective privacy~\cite{DBLP:conf/www/SquicciariniSP09}, networked privacy~\cite{boyd2012networked} and multiple-subject privacy~\cite{DBLP:conf/apf/GnesiMMMPV14}, among others. For a comprehensive overview, we refer the interested reader to \cite{DBLP:journals/csur/HumbertTH20}.
Interdependent privacy affects different types of data and data-sharing scenarios. A subset of attributes from the profile of a social network user may be harvested~\cite{DBLP:conf/fc/BiczokC13}. The location privacy of certain individuals may be threatened by sharing co-location information~\cite{DBLP:journals/tmc/OlteanuHSHH17}. Photo sharing may affect the privacy of friends and bystanders captured in the photo~\cite{DBLP:conf/ndss/OlteanuHDH18}. Even the genetic profile of an individual and associated inferrable medical information might get exposed by an eager relative (i.e., kin genomic privacy)~\cite{DBLP:conf/ccs/HumbertAHT13}. A common trait among the aforementioned scenarios is that all of them could be instantiated through a variety of third-party apps.

General privacy considerations regarding third-party apps, platforms, permissions, and ecosystems have been a strong focus area of researchers in the last decade. We do not even attempt to give a comprehensive overview here; rather, we highlight a few studies with close relations to this paper. Wang et al. studied the data collection practices of Facebook third-party apps and proposed control mechanisms that can increase transparency~\cite{wang2011third}. King et al. conducted an exploratory survey on how Facebook users interact with apps and how much they understand the privacy implications of such interaction~\cite{king2011privacy}. Androidleaks uncovered how sensitive data was used once the user gave the required permissions and the Android app was installed~\cite{gibler2012androidleaks}. Chia et al. studied app permissions, privacy risk signals, and community ratings on multiple app platforms~\cite{DBLP:conf/www/ChiaYA12}. FlowDroid and its follow-up works provided taint analysis for Android apps that shed light on potential unintended and malicious data leaks~\cite{arzt2014flowdroid}. Reardon et al. explored the many ways apps can circumvent the Android permission system~\cite{reardon201950}. Finally, Kelley et al. (and many others building on their study) showed that users actually factor in their privacy concerns when choosing between apps if they are presented with easy-to-understand privacy facts before installation~\cite{kelley2013privacy}. The above selection of studies clearly demonstrates that i) permission models are imperfect, ii) various privacy leaks do occur in apps, and iii) users act on their concerns when presented with tractable information on app privacy.

Yet, there are only a handful of scientific studies dealing explicitly with interdependent privacy situations regarding third-party apps. Biczok and Chia showed that the personal, relational, and spatial privacy of Facebook users was threatened by their friends~\cite{DBLP:conf/fc/BiczokC13}. Pu and Grossklags investigated the effect of selfish and other-regarding preferences in social app adoption~\cite{DBLP:journals/popets/PuG16}. Harkous and Aberer analyzed Google Drive apps and pointed out that users suffered more privacy loss owing to their collaborators than their own actions~\cite{DBLP:journals/corr/HarkousA17}. Finally, Symeonidis et al. presented a comprehensive data analytics, modeling, and legal study on the \emph{collateral information collection} practices of Facebook apps affecting the friends of the user~\cite{DBLP:journals/compsec/SymeonidisBSPSP18}. While these studies and further anecdotal evidence suggest that interdependent privacy issues might be the norm rather than the exception on most third-party app platforms, the research community lacks a data-driven study for available, active, but previously uncharted platforms, such as Android, browser extensions, cloud services, and videoconferencing. One of the objectives of this paper is to fill this gap.

%----added related work for IDP mitigation
%\color{ForestGreen}

There have been some studies on the sharing of information affecting the privacy of multiple people; these studies have introduced different ideas from various domains, such as law, social sciences, and information technology. Most of these ideas were not designed to solve (partially or fully) IDP issues involving third-party applications; however, they pointed out the inherent challenges around this scene.

On the legal side, pre-GDPR studies pointed out that organizations often did not have the incentives to notify users about threats or privacy breaches~\cite{DBLP:conf/mcis/KarydaM16}. Later, partly as an answer to this, new data protection regulations followed, e.g., the General Data Protection Regulation (GDPR) in Europe, the California Consumer Privacy Act (CCPA) in California, or even the Personal Information Protection Law (PIPL) in China. These regulations laid down the rights of data subjects, the obligations of organizations processing personal data, and the steps of proper information gathering and management. These groundbreaking privacy regulations increased the privacy appetite of users, but also of governments. In fact, the recent decision of the European Data Protection Board on needing explicit consent for targeted ads on Facebook and Instagram (both owned by Meta) is perceived by many as a strong step in the right direction\footnote{https://www.euractiv.com/section/data-privacy/news/eu-squeezes-meta-on-personal-data-use-for-targeting-ads/}. 
The formulation of regulations makes service providers pay more attention to the rights of the people whose information is directly collected; however, it ignores the fact that data collected from end-users may include other people's private information. Thus, owing to the complex scope of IDP and the difficulty of formulating preventive clauses, data protection laws have not covered IDP explicitly, making it somewhat of a gray zone~\cite{van2016regulating,DBLP:journals/compsec/SymeonidisBSPSP18}.

Without explicit mandatory regulations, the mitigation of IDP risks can only be implemented spontaneously, including both users and service providers. Kamleitner et al. proposed a new framework to better understand the IDP problem and outlined some principles for solutions~\cite{kamleitner2019your}. Besides some potential high-level intervention mechanisms, the article also points out the social responsibility of data controllers and processors in protecting the privacy of consumers in such a complex information-gathering scenario.
 
Few studies have articulated technical approaches. Gnesi et al. proposed to let users create privacy policies locally and standardize the storage of IDP information~\cite{gnesi2014my}. Olteanu et al. designed ConsenShare, a system for users to register personal information in and upload the photos to be shared first; users can then only share photos after obtaining the consent of other people in the photos~\cite{DBLP:conf/ndss/OlteanuHDH18}. The ``others'' indicated above must also be registered in the system and obtain identifiers. 
Both of these technology-focused solutions can be interpreted as new third-party applications. On the contrary, we aim to mitigate IDP-related risks in all third-party app platforms.

Marsch et al. proposed a design for an interdependent privacy feedback system that leverages natural language generation to communicate the potential privacy impacts of app permissions regarding both the user and others. The authors concluded that increasing awareness about the interdependent privacy consequences of app permissions could lead to more responsible privacy behavior and a more privacy-friendly app ecosystem~\cite{10.1145/3479581}.
In fact, a well-designed privacy dashboard has proved to be an efficient and user-friendly transparency-enhancing tool (TET)~\cite{DBLP:conf/primelife/RaschkeKDK17}. A privacy dashboard is a feature in apps that allows users to manage their personal data and privacy settings. Such a dashboard has also been proposed in the context of Facebook apps, specifically for raising awareness on IDP~\cite{DBLP:journals/compsec/SymeonidisBSPSP18}. In recent years, big tech companies have made a push towards more transparent data practices and providing users with greater control over their personal data\footnote{https://myaccount.google.com/dashboard} \footnote{https://legal.yahoo.com/us/en/yahoo/privacy/dashboard/index.html}, even implementing privacy shortcuts\footnote{https://www.facebook.com/help/ipad-app/395495000532167}, allowing users to quickly adjust their settings from within the app without having to navigate to a separate dashboard. We believe that user feedback on privacy should be a part of any practical risk reduction mechanism; yet, it is satisfactory in itself.
\color{black}
\section{Platforms, permissions and interdependent privacy}
\label{sec:permissions}

\subsection{Permissions and interdependent privacy}

\subsubsection{Permission-based access. }
Third-party app platforms share a common security model, which is based on requesting and granting permissions. App permissions guard the access to i) restricted data, such as location or contact information, and ii) restricted actions, such as taking photos or connecting to the Internet. Generally, the main objectives of app permissions include: i) enabling user control over data shared, ii) achieving transparency so that the user understands what data an app is using and why, and iii) promoting data minimization so that the app accesses and utilizes only the data absolutely required for a specific task the user invokes.

Platforms, e.g., Google's Android, have evolved significantly since their inception to achieve these objectives. Android has introduced install-time and run-time permissions; the latter group includes all individual permissions deemed \emph{dangerous} by the platform. Run-time permissions can be explicitly granted (or denied) by the user through a dedicated pop-up window, shown when the execution of the app reaches a state where the permission is required. On top of this, very recently, Android has included a \emph{privacy dashboard} that shows which apps had used sensitive permissions and for how long in the last 24 hours; also, with easy access to revoke said permissions if so desired. Despite all these improvements in Android and other permission mechanisms, there are still no specific (neither transparency nor mitigation) measures targeted at interdependent privacy.

Rubbing salt into the wound, app platforms' definitions of certain permissions are vague as to what extent the app will obtain and use sensitive private information. Combining this more general transparency issue with the specific flaws mentioned above, two sub-optimal privacy outcomes emerge. First, the user does not have sufficient knowledge of the scope of the information to be shared: others' private data might be transferred to the app without even their knowledge. Second, the user might grant excessive permissions to the app to preserve full functionality. Although this latter has been shown to be an issue with respect to one's own sensitive attributes, it could induce an even more negative impact in the context of interdependent privacy.

\subsubsection{Permissions related to interdependent privacy. }
Corresponding to the above two points, when the permission involved is ambiguous, users pay more attention to protecting their own privacy while ignoring the privacy of their friends~\cite{DBLP:journals/popets/PuG16}. When the number of permissions granted by users to apps becomes larger, interdependent privacy issues often emerge. An obvious example is a top-rated Firefox extension called \emph{AdBlocker Ultimate}. The permission-related warnings of this app are the following: W1) ``Access browser tabs'', W2) ``Store unlimited amount of client-side data'', W3) ``Access browser activity during navigation'', and W4) ``Access your data for all websites''. Plausibly, the combination of W1, W3, and W4 enables the extension to read the website, detect ads, and replace them with blank boxes. However, the same permissions enable the app to collect, e.g., messages sent to and received from a web-based chat; an outcome that could cause privacy loss to the communication partners of the user, an obvious interdependent privacy scenario, no user would prefer to experience. Furthermore, W2 enables the storage of unlimited personal data collected through W1, W2, and W4; this can allow for observing, e.g., personal communications over a longer period of time. Yet, not granting these requested permissions makes it impossible to install and use the app.
%Note that there is no indication of separate read and write permissions.

As we would like to quantify the extent to which interdependent privacy issues are present in third-party app platforms, we classify permissions into three pre-defined categories:
invoking interdependent privacy (IDP), potentially invoking interdependent privacy (PIDP), and not invoking interdependent privacy (NIDP). If a permission \emph{directly} enables access to private data related to a natural person other than the user herself, it is in IDP; e.g., the \verb+READ_CONTACTS+ permission in Android. If a permission \emph{potentially} enables access to private data related to a natural person other than the user herself, it is in PIDP. Such risk can be realized through i) accessing data that \emph{may} implicate  multiple parties, such as photos or documents (e.g., \verb+READ_EXTERNAL_STORAGE+ in Android); ii) enabling a restricted action that \emph{may create} multi-party data, such as photos or audio recordings (e.g., \verb+RECORD_AUDIO+ in Android); and iii) enabling \emph{inference} of other's private data with reasonable efforts, such as location via co-location information from other sources (e.g., \verb+ACCESS_FINE_LOCATION+ in Android). Note that granting a PIDP permission does not automatically constitute privacy loss for a third party; the loss is context-dependent and may require additional effort from the app developer or an adversary. If a given permission belongs neither to IDP nor PIDP, then it is in NIDP and not in our focus.

\subsection{Platform specifics}

Here, we briefly introduce the app platforms we investigated. For practical data availability reasons, we targeted the most popular mobile app platform, Android, two well-known browsers providing an API for third-party extensions (Mozilla Firefox and Opera), and Google Workspace, a cloud-based enterprise collaboration tool bundle. Although these four platforms vary greatly in both their functionality and technical mechanisms, all of them offer the equivalent of an app store, where the access control of apps is based on the user granting permissions.

\subsubsection{Android. }
Android users can download and install more than 3 million apps from the Google Play store, making Android the largest third-party app platform, both by user base and the set of available apps. This popularity has made the platform's permission model change continuously over time while trying to keep a balance between being appealing to both users and third-party developers alike. The current stable OS is v11 (with v12 right around the corner), while the API version, also defining the current permission model, is level 30. Android has evolved into a general-purpose OS with plenty of protected data objects and actions; this amounts to 91 permissions in total, offered to third-party apps). We make 91 our baseline for the total number of relevant permissions. Out of these 91, there are 4 that explicitly and 16 that potentially interfere with others' personal data instantiating interdependent privacy, see~ Table \ref{tab:googleplay}. Note that the pop-up messages, appearing when installing an app from Google Play, contain warnings that can be mapped directly to API-level permissions with reasonable effort.

% Cellular data settings
\begin{table}[tb]
\centering
\caption{Android permissions: IDP and PIDP}\label{tab:googleplay}
\begin{tabular}{ |p{6cm}|p{6cm}|  }
% \hline
% \multicolumn{2}{|c|}{Google play store permission } \\
 \hline
 IDP &PIDP  \\
 \hline
 \hline
read call log\_Phone    &  read the contents of your USB storage\_Photos/Media/Files
         \\
    \hline
read your contacts\_Contacts    &  modify or delete the contents of your USB storage\_Photos/Media/Files        \\
    \hline
modify your contacts\_Contacts    &  approx. location (network-based)\_Location         \\
    \hline
read your text messages (SMS or MMS)\_SMS
    &  precise location (GPS and network-based)\_Location             \\
    \hline
    &  access extra location provider commands\_Location            \\
    \hline
    &  take pictures and videos\_Camera          \\
    \hline
    &  read sensitive log data\_Device \& app history
          \\
    \hline
        & read your Web bookmarks and history\_Device \& app history             \\
    \hline
        &  record audio\_Microphone            \\
    \hline
        &  read the contents of your USB storage\_Storage            \\
    \hline
        &  modify or delete the contents of your USB storage\_Storage
            \\
    \hline
        &  find accounts on the device\_Contacts
            \\
    \hline
            &  read cal events plus confidential information\_Calendar
            \\
    \hline
            &  add or modify cal events and send emails to guests w/o owners' knowledge\_Calendar
            \\
    \hline
            &  read cell broadcast messages\_SMS
            \\
    \hline
            &  find accounts on the device\_Contacts
            \\
    \hline
\end{tabular}
\end{table}

\subsubsection{Browser extensions: permissions and warnings. }
Although referred to differently, browser extensions are very similar to apps. Extensions usually expand browser functionality and manage user operations. Owing to their objectives and architecture, browser extensions are all about interacting with their respective platforms, oftentimes resulting in obtaining large amounts of information about user operations in the browser in real-time, but also about content downloaded by the browser. Note that browsers are also used to access intranets and other non-public resources, therefore, they might leak a variety of personal (and other confidential) information if something goes wrong. Both Firefox and  Opera are based on Chromium, therefore their APIs and permission models facing third-party extensions are all based on the Chrome API (along with Chrome, Edge, Brave and Safari, to be correct). Both browsers have their own extension store. 

Albeit they are based on the same APIs, Firefox and Opera have some unique characteristics. They both support the majority of permissions but not all\footnote{\url{https://developer.mozilla.org/en-US/docs/Mozilla/Add-ons/WebExtensions/manifest.json/permissions#browser_compatibility}}, and they both define their own warning messages that users can see before/when they install an extension\footnote{e.g. Firefox: \url{https://support.mozilla.org/en-US/kb/permission-request-messages-firefox-extensions}}. In fact, Opera does not show these warnings when installing; they are only visible on their dedicated page in the extension store. Making things more complicated, i) not all permission requests generate warning messages, and ii) warning messages and API-level permissions are not totally consistent: the platforms have decided to simplify warnings for the sake of clarity to the average user. While this is laudable from one aspect, these explanations sometimes do not fully reflect the risks of granting the requested permissions. Exact mappings between permissions and user warnings are hard to find but may be extrapolated from Chrome's official documentation\footnote{\url{https://developer.chrome.com/docs/extensions/mv2/permission_warnings/#permissions_with_warnings}}. Since it is only feasible to scrape the extension stores for per app warnings (and not for API-level permissions), we base our analysis on these. Note that our datasets contain information on Manifest V2 extensions; however, the changes introduced in Manifest V3 do not have a significant impact on interdependent privacy\footnote{\url{https://developer.chrome.com/docs/extensions/mv3/intro/mv3-overview/}}.

Firefox has 26 different warning messages, 19 of which are potential culprits for interdependent privacy violations, see Table \ref{tab:browsers}. Opera extensions make use of 20 types of warning messages, 13 of which pose a potential threat owing to privacy interdependence, see Table \ref{tab:browsers}. 
%Finally, the Microsoft Edge platform utilizes 10 different warning message, 6 of which indicate a potential interdependent privacy breach, see Table \ref{tab:edge}. 
Note that all affected warnings (and their corresponding permissions) are in PIDP, and we omit NIDP due to space constraints. Also, note that all Opera warnings start with ``This extension (can/will)''.

\begin{table}[tb]
\centering
\caption{Browser extension permissions: PIDP}
\label{tab:browsers}
\begin{tabular}{ |p{6cm}|p{6cm}|  }
 \hline
 %\multicolumn{2}{|c|}{Firefox permission List} \\
 %\hline
 Firefox & Opera \\
 \hline
 \hline
    
     Access browser tabs & Access your data on all websites \\
    \hline
     Access browser activity during navigation & Access your tabs and browsing activity \\
    \hline
     Access your data for named site & Access your data on some websites\\
    \hline
     Exchange messages with programs other than Firefox & Exchange messages with programs other than Opera\\
    \hline
     Download files and read and modify the browser's download history & Capture the content of the entire screen or of individual tabs and windows \\
    \hline
     Access your location & Access data you copy and paste \\
    \hline
     Access recently closed tabs &  Allow other installed extensions and web pages to communicate with this extension \\
    \hline
     Access your data for all websites & Detect your physical location \\
    \hline
     Store unlimited amount of client-side data & Manipulate privacy-related settings \\
 \hline
      Access your data for sites in the named domain & Know which sites you're visiting most often \\
    \hline
     Read and modify bookmarks & Read and modify bookmarks \\
    \hline
      Access your data on \# other sites & Store an unlimited amount of client-side data \\
    \hline
     Get data from the clipboard & Read and modify your browsing history \\
    \hline
    Extend developer tools to access your data in open tabs & \\
    \hline
     Read the text of all open tabs & \\
    \hline
     Access browsing history & \\
    \hline
     Access your data in \# other domains & \\
    \hline
     Access browsing history & \\
    \hline
    Read and modify browser settings & \\
    \hline
\end{tabular}
\end{table}

\subsubsection{Google Workspace (formerly GSuite). }
GSuite is a collaborative enterprise office platform launched by Google in September 2016 that was rebranded to Google Workspace in 2020, when it already had more than 2 billion users. Users do not need to download applications, they only need to edit and share files in the cloud, realizing remote collaboration. The platform has an app store, the Marketplace\footnote{\url{https://workspace.google.com/marketplace}}, where business, productivity, and educational tools are offered by third-party developers. One of Workspace's subsystems, Google Drive, has already been shown to leak others' personal information through apps owing to its collaborative nature~\cite{DBLP:journals/corr/HarkousA17}; however, its permission model has changed completely due to the integration of various Google subsystems into the Workspace. The platform has many specialized permissions catering to its intended usage as a collaborative office productivity solution. Specifically, there are 87 different permissions, out of which 3 are IDP (``See and download your contacts'', ``View customer-related information'' and ``View, edit, or permanently delete contacts'' and 71 are PIDP (which we omit due to the lack of space). Note that although Workspace is a subscription-based service for enterprises and universities, it hosts huge amounts of private data. What is more, if an employee (usually a system administrator) installs a third-party app, resulting in a privacy violation for other natural persons, the company can be held responsible as per the GDPR.

\subsubsection{Zoom Marketplace. }
%\color{ForestGreen}
\label{sec:Zoom}
In recent years, the COVID-19 pandemic has prompted the emergence and popularity of novel tools and platforms, with some becoming indispensable in daily life, such as \emph{Zoom}. Owing to its exceptional teleconferencing and remote collaboration capabilities, Zoom continues to thrive even after the lockdown period. Zoom Meeting, a cross-platform application for Windows, macOS, iOS, Android, Chrome OS, and Linux, is recognized for its user-friendly interface and functionality. Key features encompass one-on-one meetings, group video conferences, screen sharing, plugins, browser extensions, and the ability to record meetings and automatically transcribe them.

Zoom Marketplace is the third-part app platform of Zoom; it employs a permission-based system, requiring users to activate a pre-approval switch before installing an application. This switch is unrelated to the permissions requested by the app. Once the switch is enabled and the ``add'' button is clicked, specific permissions and authorization options are displayed. A less prominent authorization option above the ``authorize'' button is unrelated to the app's installation. This option, termed ``shared access permissions'', lacks a detailed description of its scope and usage. According to Zoom Meeting, this permission means that the app will perform some actions on the user's behalf, such as scheduling a meeting, for which the app will have to obtain this permission and have access to some of the user's information, such as user profiles, and the information of other users of the account.\footnote{\url{https://support.zoom.com/hc/en/article?id=zm_kb&sysparm_article=KB0063819#h_01FMWTFQ9RM20TQB9HANS05AMQ}}

\begin{figure}[tb]
\centering
\includegraphics[width=1.0\textwidth]{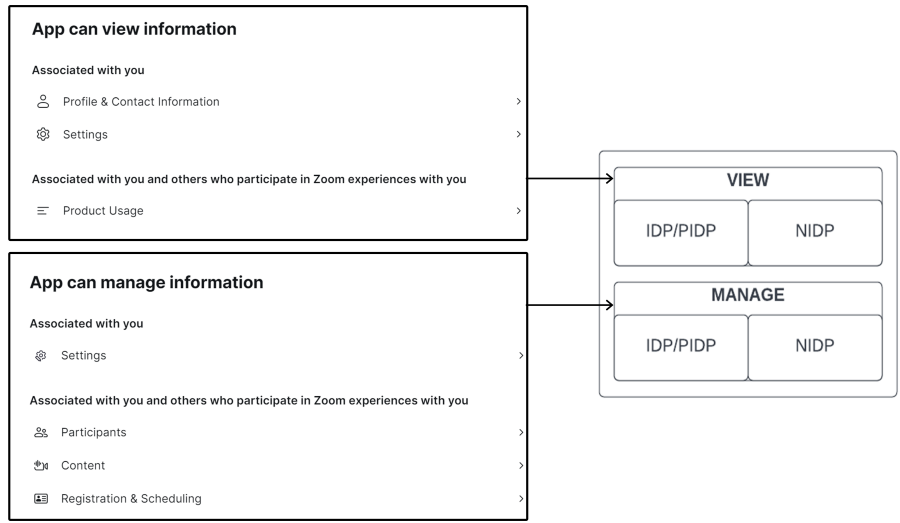}
\vspace{-3mm}
\caption{Zoom Marketplace permission system structure} 
\label{fig:Zoomstruc}
\vspace{1mm}
\end{figure}

The authorization option is the critical determinant; users must either accept all permissions or not use the app. The various permissions the app will acquire are detailed on the page. Zoom Marketplace is the first platform found to attempt to differentiate between IDP and NIDP in the permission system. Zoom categorizes permissions into ``App can view information'' and ``App can manage information''. Upon reviewing the API documentation, it was discovered that each subset contains two types of instructions. Permissions are divided into two subsets: ``Associated with you'' and ``Associated with you and others you're allowed to access''. The second set of descriptions consists of: ``Associated with you, others you're allowed to access, and others included in that information'' and ``Associated with you and others who participate in Zoom experiences with you''. Comparing the API documentation with the content displayed on the webpage reveals that for each category on the webpage, the subset displays only one description. The two descriptions have an inclusive relationship, where ``Associated with you and others you're allowed to access'' contains ``Associated with you'' and ``Associated with you and others who participate in Zoom experiences with you''. Interestingly, this classification system (see Fig. \ref{fig:Zoomstruc}) essentially adheres to the IDP concept\ref{fig:Zoomstruc}, horizontally partitioning permissions related solely to the user or those the user can control into one category and those the user cannot control and involve others into another category. Vertically, permissions are divided into view and manage categories. This approach provides more granular control of permissions and enhances transparency and user awareness on the platform. Nevertheless, IDP/PIDP permissions are still present on the platform (see Tables \ref{tab:zoomview} and \ref{tab:zoommanage}).

\begin{table}[tb]
\centering
\caption{Zoom: View permissions}\label{tab:zoomview}
\begin{tabular}{ |p{6cm}|p{6cm}|  }
% \hline
% \multicolumn{2}{|c|}{Google play store permission } \\
 \hline
 IDP\&PIDP & NIDP  \\
 \hline
 \hline
Profile \& Contact Information    &  Settings
         \\
    \hline
Product Usage    &  Registration Information        \\
    \hline
Device Information    &  Account Information         \\
    \hline
Participant Profile \& Contact Information 
    &               \\
    \hline
Content     &              \\
    \hline
Calendars   &            \\
    \hline

\end{tabular}
\end{table}

% \begin{table}[tb]
% \centering
% \caption{Zoom: View permissions}\label{tab:zoomview}
% \begin{tabular}{ |p{6cm}|p{6cm}|  }
% % \hline
% % \multicolumn{2}{|c|}{Google play store permission } \\
%  \hline
%  NIDP & IDP\&PIDP  \\
%  \hline
%  \hline
% Settings    &  Registration Information
%          \\
%     \hline
% Profile \& Contact Information    &  Product Usage        \\
%     \hline
% Content    &  Content         \\
%     \hline
% Product Usage
%     &  Participant Profile \& Contact Information             \\
%     \hline
% Account Information    &              \\
%     \hline
% Calendars   &            \\
%     \hline
% Device Information    &  
%            \\
%     \hline

% \end{tabular}
% \end{table}

\begin{table}[tb]
\centering
\caption{Zoom: Manage permissions}\label{tab:zoommanage}
\begin{tabular}{ |p{6cm}|p{6cm}|  }
% \hline
% \multicolumn{2}{|c|}{Google play store permission } \\
 \hline
IDP\&PIDP & NIDP  \\
 \hline
 \hline
Profile \& Contact Information    &  Settings
         \\
    \hline
Registration \& Scheduling    &  Account Information       \\
    \hline
Participants    &           \\
    \hline
Content
    &              \\
    \hline
Devices
    &              \\
    \hline

\end{tabular}
\end{table}

% \begin{table}[tb]
% \centering
% \caption{Zoom: Manage permissions}\label{tab:zoommanage}
% \begin{tabular}{ |p{6cm}|p{6cm}|  }
% % \hline
% % \multicolumn{2}{|c|}{Google play store permission } \\
%  \hline
%  NIDP & IDP\&PIDP  \\
%  \hline
%  \hline
% Profile \& Contact Information    &  Registration \& Scheduling
%          \\
%     \hline
% Settings    &  Content       \\
%     \hline
% Account Information    &  Participants         \\
%     \hline

%     &  Devices            \\
%     \hline

% \end{tabular}
% \end{table}

It is straightforward to see that each platform has a significant proportion of its permissions and warnings connected to interdependent privacy; see Table \ref{tab1}. This answers \emph{RQ1} affirmatively: \emph{interdependent privacy issues are indeed pervasive among third-party app platforms. }Note that although new versions of platforms are rolled out periodically, the IDP issues are unlikely to disappear: they are inherently present owing to permission-based access control. Unless the access control mechanisms of app platforms are completely re-designed, interdependent privacy is here to stay.

\begin{table}[tb]
\centering
\caption{Summary: IDP and PIDP permissions/warnings}
\label{tab1}
\begin{tabular}{ %|p{3cm}||p{3cm}|p{3cm}|p{3cm}|  }
|p{3cm}||c|c|c|  }
 \hline
% \multicolumn{4}{|c|}{Permission List} \\
 %\hline
 %\hline
 Platform & No. of permissions/warnings & IDP + PIDP & Ratio\\
 \hline
 \hline
 Android & $91$ & $20$ & $21.98\%$\\
 \hline
 Firefox &  $26$  & $19$ & $73.08\%$\\
 Opera & $20$   & $13$ &   $65\%$ \\
 %Edge &  10  & 6   &60\%\\
 \hline
 Google Workspace  & $87$ & $74$ &  $85.06\%$\\
 Zoom Marketplace  & $12$ & $9$ &  $75\%$\\
 \hline
\end{tabular}
\end{table}

\color{black}
\section{Application-level statistics}
\label{sec:data}

\subsection{Data collection}
We collected datasets by scraping the app stores of 5 different third-party app platforms: Android ($10,589$ apps), Mozilla ($16,546$), Opera ($1,682$), and Google Workspace ($882$) in late 2020 and early 2021; and Zoom App Marketplace ($1,841$) in early 2023. Each record contains all available meta-data, e.g., app name, category, permissions/warnings, number of users, rating, etc., depending on the actual platforms. Due to the app stores' protection against scraping, i) we did not manage to collect enough data for Chrome and Edge extensions; therefore, we omitted these platforms from our analysis; ii) our Android dataset contains only a fragment of the millions of available apps (yet, large and random enough to be significant). To the best of our knowledge, we collected complete datasets (i.e., reachable by scraping) for Firefox, Opera, Google Workspace, and Zoom App Marketplace. Note that automatic scraping was infeasible for Google Workspace; we manually collected information on all available apps. 
% \color{ForestGreen}
% We also scraped data directly from the website of  Zoom App Marketplace\footnote{\url{https://marketplace.zoom.us/}}. 
By collecting URLs of different applications, we could track detailed information, including names, categories, and permissions requested by each app. Note that Zoom Marketplace does not list features like the number of users and ratings, making it hard to measure the popularity of Zoom apps. All datasets and scraping scripts are available for download.\footnote{\url{https://github.com/shuai20/IDP_Filter}}
\color{black}
\subsection{Do real apps request IDP/PIDP permissions?}

Table \ref{tab:ippip} shows the number of apps that requested at least one IDP or PIDP permission. The last column calculates the proportion of the union of these apps versus all the apps in the dataset. It is clear that the vast majority of apps are affected, as evidenced by a proportion larger than $80\%$ for all platforms. Note that the browser platforms offer only PIDP permissions.
 
\begin{table}[tb]
\caption{Number of apps with IDP/PIDP}
\label{tab:ippip}
\centering
\begin{tabular}{ |p{2.8cm}||c|c|c|c|  }
 \hline
% \multicolumn{4}{|c|}{Google play store permission } \\
% \hline
 Platform& Apps with IDP & Apps with PIDP & Total Apps & Ratio\\
 \hline
 \hline
Android &  $1,029$   &  $8,307$   & $10,589$ & $78.66\%$  \\
    \hline
Firefox &   0   & $13,704$   & $16,546$  & $82.82\%$ \\
    %\hline
Opera & 0  &  $1,421$   &  $1,682$   & $84.48\%$   \\
    \hline
Google Workspace &   $29$   & $845$   & $882$  & $97.62\%$ \\
Zoom Marketplace &   $1,832$   & $1,803$   & $1,841$  & $99.51\%$ \\
    \hline

 \hline
\end{tabular}
\vspace{-5mm}
\end{table}
 
 To further study the privacy protection permissions of apps, we calculated the permissions requested by each app. 
 Regarding Android (Google Play store), $17.2\%$ of apps requested the permission ``Contacts'', which means that their users have shared their contact list with the third-party developer, directly exposing others' personal data without their knowledge. Besides, $78.66\%$ of apps have the potential to leak private information owing to interdependent privacy. On average, each app requests $11.21$ permissions, out of which $4.4$ are IDP or PIDP.
 
 Mozilla and Opera extensions, despite their similar architecture, differ significantly in terms of PIDP warning types ($0.83$ vs. $3.93$ on average) and total warning types ($0.85$ vs. $4.63$) displayed. Note that, although here we observe warning instead of permissions, the difference holds, as warning-permission mappings are alike on both platforms. One reason could be that more Mozilla extensions make use of the \verb+active_tab+ permission (which does not generate a warning) instead of \verb+_url+ type permissions\footnote{\url{https://developer.mozilla.org/en-US/docs/Mozilla/Add-ons/WebExtensions/manifest.json/permissions#activetab_permission}}. Some other permissions also do not generate warnings, therefore the total number of permissions requested in Table \ref{tab:pipapp} is underestimated, while the proportion in the last column is overestimated for browser extensions.

 Google Workspace is dedicated to collaborative enterprise features with a lot of PIDP permissions therefore we expected a high proportion of those requested by apps. Indeed, $85\%$ of total permissions requested ($2.04$ out of $2.42$) are IDP or PIDP. We also observed that a majority of permissions are requested only by a few apps. This can be explained by the relatively low number of available apps and the fact that permissions are very specific (especially compared to browser extensions), e.g., ``View and manage your Google Slides presentations'' instead of ``View and manage your documents''.
 
 %\color{ForestGreen}
 There are a total of $2,433$ apps on Zoom Marketplace, belonging to 32 different categories. On average, each app has $2.54$ vNIDP (view NIDP) permissions, $2.7$ vIDP/vPIDP permissions, $0.25$ mNIDP (manage NIDP) permissions, and $2.99$ mIDP/mPIDP permissions. From a statistical point of view, the number of IDP/PIDP permissions is higher, especially among manage permissions. It is generally believed that the permission level of ``app can manage'' is higher than that of ``app can view'', and the risk is also greater.
\color{Black}
Based on the results above, we can also answer \emph{RQ2} affirmatively: \emph{actual apps do request IDP/PIDP permissions enabling collateral information collection on all studied platforms.} 
%\color{ForestGreen}
While new platform versions and newly released apps might change the absolute number of IDP/PIDP permissions per app, their magnitude is unlikely to change (pending no complete re-design of the respective permission structures).

\begin{table}[tb]
\centering
\caption{Average number of IDP\&PIDP permissions per app}
\label{tab:pipapp}
\begin{tabular}{ |p{2.8cm}||c|c|c|c|  }
 \hline
% \multicolumn{4}{|c|}{Google play store permission } \\
% \hline
 Platform & IDP/PIDP permissions & Total permissions & Proportion\\
 \hline
 \hline
Android  &  $4.40$    &  $11.21$ & $39.3\%$   \\
    \hline
Firefox    & $0.83$  & $0.85$  & $97.6\%$ \\
    %\hline
Opera &  $3.93$  &  $4.63$ & $84.9\%$   \\
    \hline
Google Workspace    & $2.04$  & $2.42$  & $84.3\%$ \\
Zoom Marketplace  & $6.20$ & $7.90$ &  $78.5\%$\\
    \hline
\end{tabular}
\end{table} 

\subsection{Risk Signals}

Users can obtain limited information when deciding upon installing third-party apps, such as category, number of users, user ratings, and permission types. Taking the Google Play store as an example, here we investigate whether the user can interpret these pieces of information as risk signals towards interdependent privacy. Previous studies found that neither popularity (number of users), nor community ratings (stars) are good indicators for privacy-conscious app behavior~\cite{DBLP:conf/www/ChiaYA12}. We also found evidence supporting this hypothesis. In fact, community ratings show a weak positive correlation with both the number of total permissions and the number of IDP/PIDP permissions requested: favorable ratings are mostly based on advanced functionality requiring more permissions.

\begin{figure}[tb]
 \centering
\includegraphics[width=0.9\textwidth]{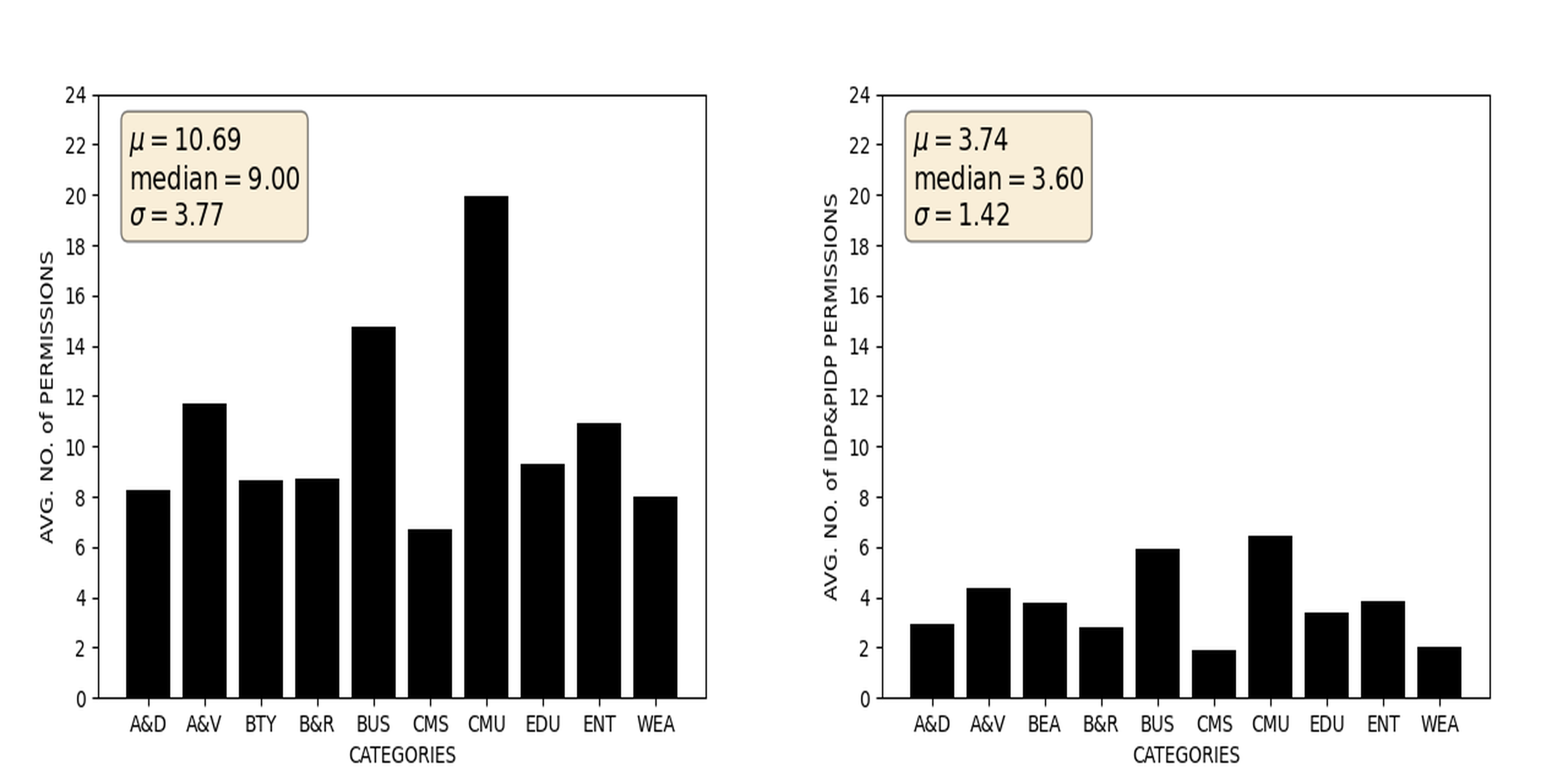}
\vspace{-3mm}
\caption{Average number of permissions per app with different categories; Average number of IDP/PIDP permissions per app with different categories. 
%"A\&D"Art \& Design "A\&V" Auto \& Vehicles, "BEA" Beauty, "B\&R" Books \& Reference, "BUS" Business, "CMS" Comics, "CMU" Communication, "EDU" Education, "ENT" Entertainment, "WEA" Weather
Art\&Design, Auto\&Vehicles,  Beauty, Books\&Reference, Business, Comics, Communication, Education, Entertainment, Weather} \label{fig:apps_number}
\vspace{-5mm}
\end{figure}

The only promising indicator for an app to enable collateral information collection was its category. In order to demonstrate this, we selected $2,043$ apps from the Google Play dataset randomly, with $\approx 200$ samples in each of the $10$ major categories. The average number of total permissions (left) and IDP/PIDP permissions (right) can be seen in Figure~\ref{fig:apps_number}.
%\ref{fig:Zoomcat}

The number of permissions varies greatly across categories. Apps belonging to ``Business'' and ``Communication'' request an average of $14.74$, $19.92$ permissions, while ``Art\&Design'' and ``Comics'' only have $8.26$, $6.71$. A reasonable explanation for this result is that communications and business apps have more advanced features requiring more permissions. Interestingly, the same result holds for IDP/PIDP permissions. Categories with a high number of total permissions have a high number of IDP/PIDP permissions, and vice versa. The reasoning above can explain this result partially, but we argue that it is a characteristic of communication/business apps to involve more collaboration and multi-party interaction, a main theme behind permissions invoking interdependent privacy.

\begin{figure}[tb]
\centering
\includegraphics[width=0.9\textwidth]{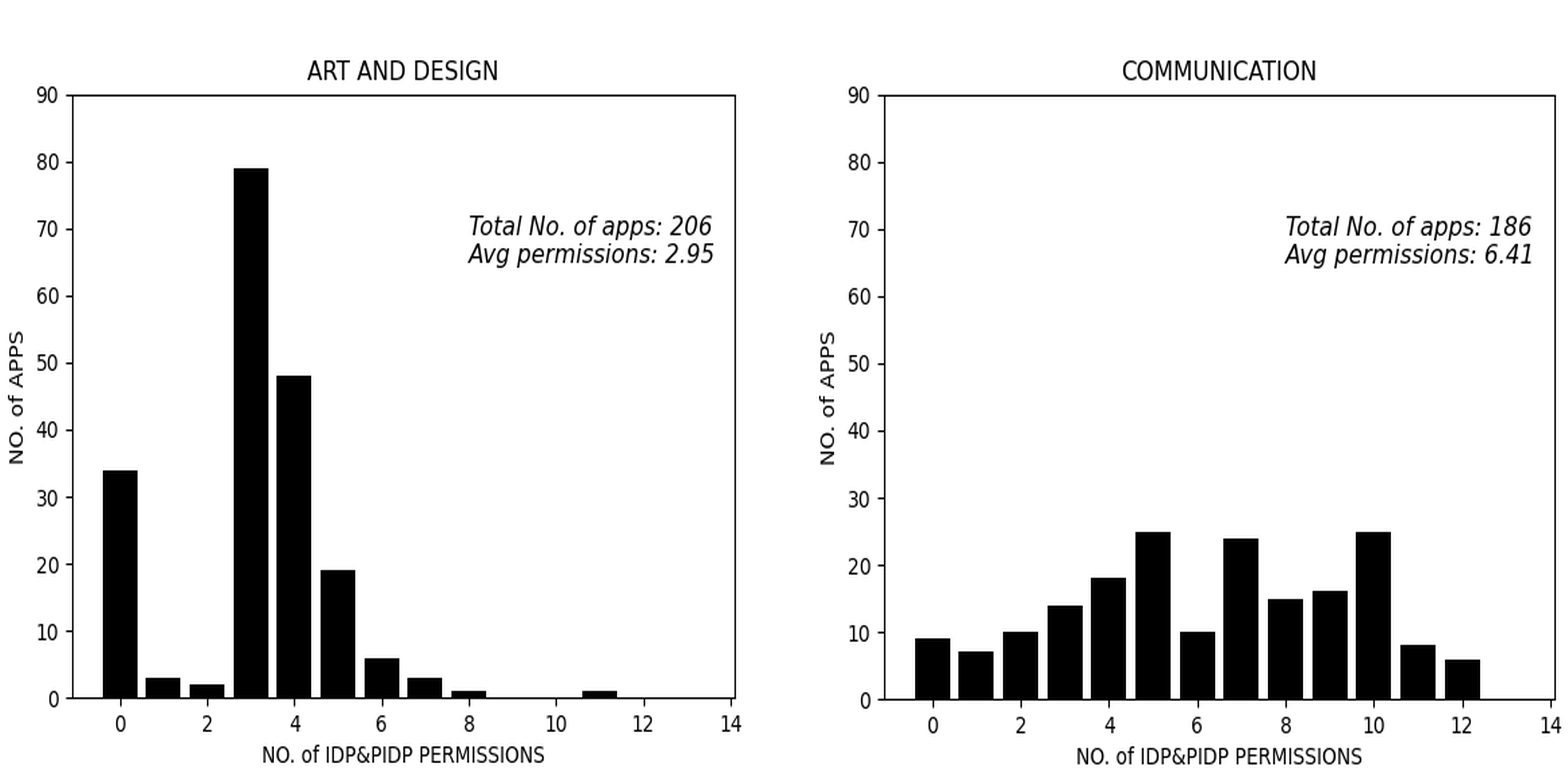}
\vspace{-3mm}
\caption{Number of apps with different number of IDP/PIDP permissions in ART\&DESIGN; Number of apps with different number of IDP/PIDP permissions in COMMUNICATION.} \label{fig:art&comm}
\vspace{-5mm}
\end{figure}

To illustrate this observation, we turn to the distribution of the number of IDP/PIDP permissions across all apps in a given category. Figure \ref{fig:art&comm} shows the histogram for this metric for the categories ``Art\&Design'' (left) and ``Communication'' (right). The difference between the two plots is striking: both the average number of IDP/PIDP permissions ($2.95$ vs. $6.41$) and the shape of the histograms (top-heavy vs. normal-like) are very different. These patterns are mostly consistent for categories with a low and high number of permissions, respectively. This corroborates our previous observation, as more interactive/collaborative categories have more apps requesting a large number of IDP/PIDP permissions.

Note that our observations on risk signals are Android-specific; more narrow-focused app platforms are unlikely to have the same risky categories, as the range of apps offered is constrained.
%It constitutes important future work for us to investigate these signals with respect to other platforms.

\section{Mitigation: considerations, objectives and design principles}
\label{sec:mitigation}
\color{black} 
\subsection{Avoidance, transparency and control}
\label{sec:discussion}

In Section \ref{sec:permissions} and \ref{sec:data}, we observed that i) all observed platforms offer permissions potentially invoking interdependent privacy, and ii) real apps do request a number of these permissions. Users are not particularly aware of interdependent privacy risks~\cite{DBLP:journals/compsec/SymeonidisBSPSP18}, and app platforms neither i) do a good job of informing the installing user and other persons affected by this issue, nor ii) offer control levers to influence such sharing. Therefore, it is up to the individual privacy awareness of installing users (acting as ``amateur data controllers''~\cite{DBLP:journals/compsec/SymeonidisBSPSP18}) and blind luck, that none of these platforms will experience its own Cambridge Analytica moment. Naturally, these options are neither satisfactory nor systemic; in the following, we discuss potential mitigation mechanisms, promoting risk avoidance, transparency, and control.

\subsubsection{Risk avoidance. }

A visceral response by app platforms to avoid exposed interdependent privacy risks could be to banish (most) IDP/PIDP permissions from their API. In fact, this is exactly what Facebook did in 2018 in response to the Cambridge Analytica scandal: it gutted its API for third-party apps and introduced strict manual app review (hiring thousands of new employees)\footnote{\url{https://about.fb.com/news/2018/04/restricting-data-access/}}. As evidenced by the declining popularity of Facebook apps, this might not be the most efficient way to deal with such risks. Indeed, the strong two-sided network effects characterizing app platforms require catering for both users and developers~\cite{parker2005two}.

\subsubsection{Transparency. }
Inspired by the GDPR and defined eloquently by Kamleitner's 3R insight framework~\cite{kamleitner2019your}, the sharing party (i.e., the amateur controller) can take three steps to reduce interdependent privacy risks: realize that there is a data transfer, recognize others' rights and respect others' rights. It is clear that transparency-enhancing technologies can facilitate the first two steps.
A potential way to make the sharer aware of interdependent privacy is to add a special warning sign to the already existing permission notification dialogues~\cite{Franz2022}. Such a solution has to be platform-specific and needs the cooperation of the platform owner. If such cooperation is unlikely, a dedicated interdependent privacy dashboard app can be implemented in the manner of proposed dashboard designs for Facebook apps~\cite{DBLP:journals/compsec/SymeonidisBSPSP18}. Note that an exact public mapping of API-level permissions to user warnings could also improve awareness (especially for browser extensions).
 
Following the opinion of the Article 29 Working Party and the subsequent recommendations of Privacy International\footnote{\url{https://privacyinternational.org/node/2997}} in the TrueCaller case, affected data subjects (i.e., ``others'') should/could also be notified by the app developer using SMS, using the very data it acquired unlawfully (i.e., contact list). Such notification, however, is not a general possibility: it depends on the platform and the actual data collected.
 
\subsubsection{Control. }
There are some privacy best practices that, when adhered to, would improve the situation on the developer and the platform owner side. These include requesting the exact minimum privileges an app needs (developer) and introducing well-defined, fine-grain permissions to enable asking for the minimum privilege (platform owner, especially for browser extensions).

Best practices aside, there is potential for interdependent privacy-specific solutions that can enable better control of personal information both for affected users, privacy-conscious apps, and platforms. Notifying affected data subjects and asking for their consent can be feasible for i) specific data types (e.g., contact lists) or closed platforms such as Facebook (where all data are connected to other users of Facebook).
In cases where a certain data object is clearly connected to multiple natural persons (e.g., photos, messages, collaborative documents, calls), sharing mechanisms tailored to multi-party data
may be utilized~\cite{such2017photo,DBLP:conf/ndss/OlteanuHDH18}. Whether these can be incorporated efficiently into a third-party app platform remains to be seen. Another way to go is to combine permissions with enforceable policies (in the manner of ~\cite{fragkaki2012modeling} but regarding privacy) and control the information flow in run-time~\cite{jia2013run} (not between components, but among platform, developer, users and others affected). An interesting restriction would be to keep data acquired through IDP/PIDP permissions locally on the user device, enabling computation (if needed for full app functionality) but restricting data transfer. There are many challenges for such a solution, starting with non-structured data that is hard to label as ``multi-party''.

Indeed, we can make a case for interdependent privacy being inherently present in current app platforms. A radical solution to mitigate this situation would be completely redesigning the currently widespread permission-based access for app platforms and trying different alternatives.

%\color{ForestGreen}
\subsection{System objectives}
\label{sec:Technoical sol}

\begin{figure}[tb]
\centering
\includegraphics[width=0.9\textwidth]{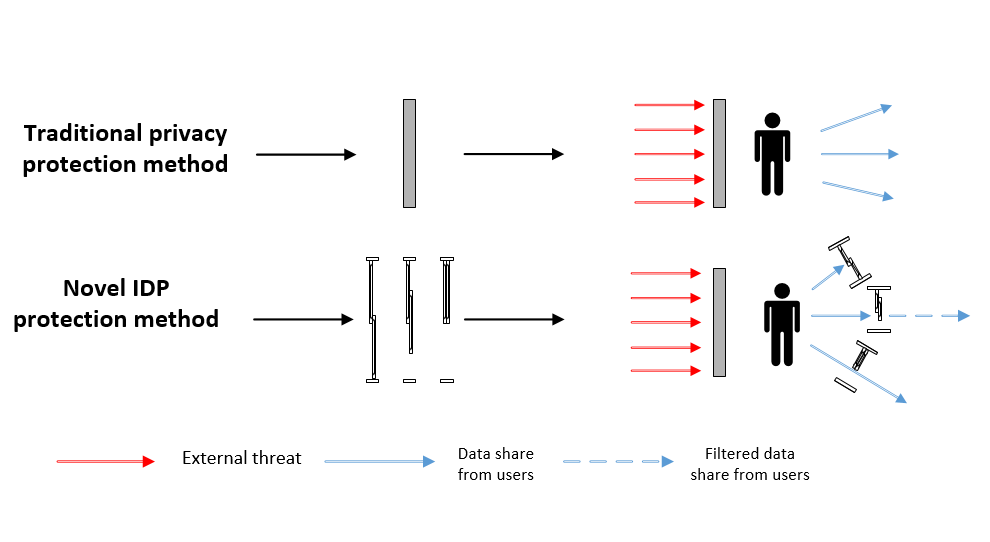}
\vspace{-3mm}
\caption{Traditional privacy protection methods vs IDP protection methods.} \label{fig:IDPvsTraditional}
\vspace{-5mm}
\end{figure}

Traditional privacy protection techniques in the third-party application domain predominantly focus on fine-tuning permissions and implementing usable privacy dashboards for users. In contrast, our emphasis regarding the mitigation of IDP issues lies in implementing an additional layer of protection during the transmission of user data when users willingly share their own information but may unknowingly share others' personal data. This ensures enhanced privacy even in instances of voluntary data sharing; see Figure \ref{fig:IDPvsTraditional}.

Designing solutions to IDP problems faces multiple challenges.
First, contemporary third-party application platforms have already achieved significant scale, making it impractical to fundamentally redesign or modify their internals solely to address IDP concerns. Moreover, the IDP issues faced by different platforms are similar but not identical. 
Second, in the absence of mandatory legal and/or institutional mandates, both app platforms and users lack incentives and enthusiasm for addressing IDP-related matters. 
%Third, given the intricacy of IDP, tackling its challenges inevitably demands a protracted duration.

Our main objectives are threefold. 
First, the system should not be tethered to any particular platform; it should be universally applicable to present-day third-party application platforms. 
Second, in light of the current absence of binding laws and regulations pertaining to IDP, the solution should be viable under the current data protection climate, offering adequate incentives for both application providers and users to adopt.
Last, under the premise of voluntary user adoption, the solution should strive to protect user IDP with high efficiency.

\subsection{Design principles}
\label{sec:design_princ}
Since no system currently exists to protect users against IDP threats on any third-party application platform, the new system should be built to adhere to specific rules to i) ensure its effectiveness and robustness and ii) minimize the likelihood of creating new privacy risks. For tractability reasons, the new system is not intended to tear down and rebuild all the permission-related mechanisms of the original platform but to enhance and complement it.

\noindent \textbf{P1: Feedback to users. } Although, in general, original permission-related warnings do state the types of data the app in question would obtain upon installation/usage, this information is usually not communicated efficiently to the user. Specifically with IDP, related risks are not even acknowledged by the platform owner or the app developer (e.g., transferring a contact list with others' contact information). Therefore, our mechanism should both i) acknowledge the IDP risk and ii) convey rich information to the user. Enriching the feedback about IDP risks helps users respect others' rights, enabling them to cater to their other-regarding preferences. As such, the IDP and PIDP classifications of the requested permissions, as well as related filtering events, should be clearly displayed in the new warnings.

\noindent \textbf{P2: Configurability. }
Data under IDP risks can be complex and of different types. Furthermore, users have different sharing preferences and requirements regarding different apps and use cases. Therefore, the system should be able to process/filter IDP-relevant information in a fine-grained manner. Suppose that the sharing of IDP-relevant information is necessary (to provide a complex functionality or better user experience. Yet, Alice may want to share her friend's location with an app but not her name, e-mail, or phone number (Scenario 1). Or, she may want to share part of her address book with a productivity app (e.g., company colleagues) but not everything (e.g., personal contacts) (Scenario 2).
Scenario 1 shows that the processing of IDP-relevant information should be fine-grained, and the system should be able to classify and process different types of information. Additionally, Scenario 2 shows that the same type of information may have different scopes and user expectations. The system should be able to treat the same type of information as a range and extract the part the user wants. Moreover, users should be able to determine which of their own data could be collected through apps installed by others.

\noindent \textbf{P3: Sufficient incentives and voluntary usage. }The new system requires additional technical means to achieve IDP protection. This affects both third-party app providers and app users. For app providers, there is a financial incentive to avoid fines imposed by data protection authorities. Given that IDP is a gray area in the GDPR and other data protection regulations, it could be wise to adopt the proposed system. Nevertheless, we propose completely voluntary participation: this signals a privacy-conscious attitude towards users and potentially results in the enhanced reputation of the participating app providers.
Note that the system should be capable of functioning without the active participation of users with a sub-optimal default configuration. However, users’ input is vital for the system to reach its potential. Users have a dual incentive: they can satisfy both their self-regarding and other-regarding privacy preferences by enabling protection against IDP issues for themselves and others, respectively.

\noindent \textbf{P4: Usability. }
It is essential that the IDP mitigation mechanism i) is easy to use (both for app providers and users), and ii) does not impede the users' quality of experience regarding normal app usage.
First, providing a mechanism that can be used on any app platform is a significant step in the right direction: this way, both app providers and users can use the same one-stop-shop IDP mitigation service across all their apps and platforms. Second, a simple interface for both app providers (e.g., a standard REST API) and users (e.g., a simple web interface for setting their preferences) facilitates easy usage. Furthermore, it is important that the service does not impose a large overhead with respect to delay to preserve the normal user experience.

\noindent \textbf{P5: Incremental deployability. }
We have already stated that we do not intend to redesign the permission systems of (popular) third-party application platforms. Although such an approach could attack the root of IDP issues, it is not practical: only platform owners (e.g., Google, Apple, Zoom, etc.) have the rights and power for such fundamental re-engineering. Instead, our IDP mitigation mechanism should operate on top of the existing app architecture, requiring zero change in the underlying mechanisms. Coupled with voluntary usage (P3), this ensures flexibility and gradual transition, where even a single app provider can opt-in to IDP protection, improving the privacy of its users instantly. This way, no critical mass (neither app providers nor users) is needed for the mitigation solution to start operating.

Note that our principles map well to the foundational principles and strategies of Privacy-by-Design~\cite{pbd,hoepman_pbds,gurses2011engineering}. The only notable absence is that of the principle of privacy embedded into the design; in fact, we aim to mitigate the lack of IDP-related protection mechanisms in third-party app platforms and the apps themselves, utilizing the exact strategies outlined in~\cite{hoepman_pbds}.

\subsection{System design}
\label{sec:system_design}

\begin{figure}[tb]
\centering
\includegraphics[width=0.9\textwidth]{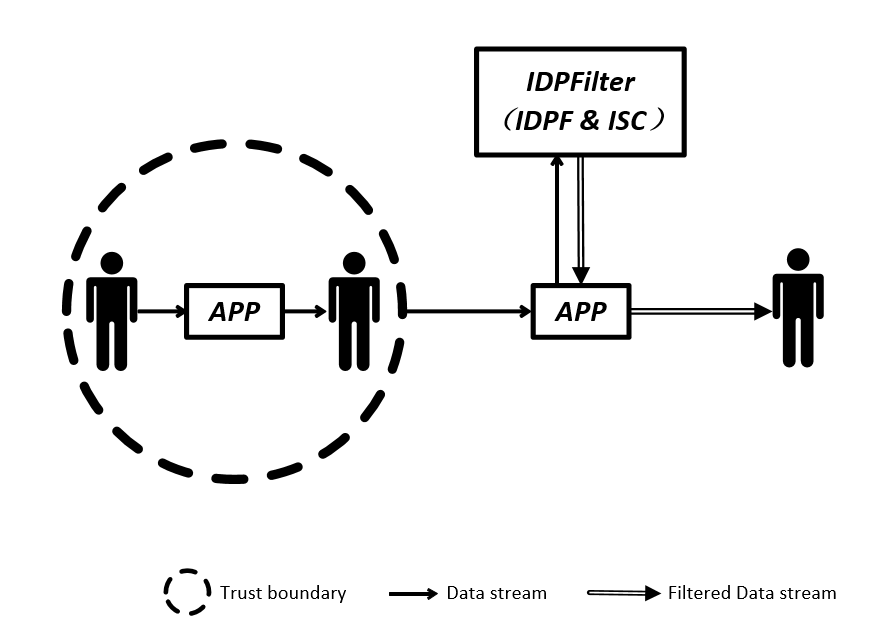}
\vspace{-3mm}
\caption{When sending information outside the trust boundary, IDP information will be filtered} 
\label{fig:User-APP-API2.0}
\vspace{1mm}
\end{figure}

\begin{figure}[tb]
\centering
\includegraphics[width=0.9\textwidth]{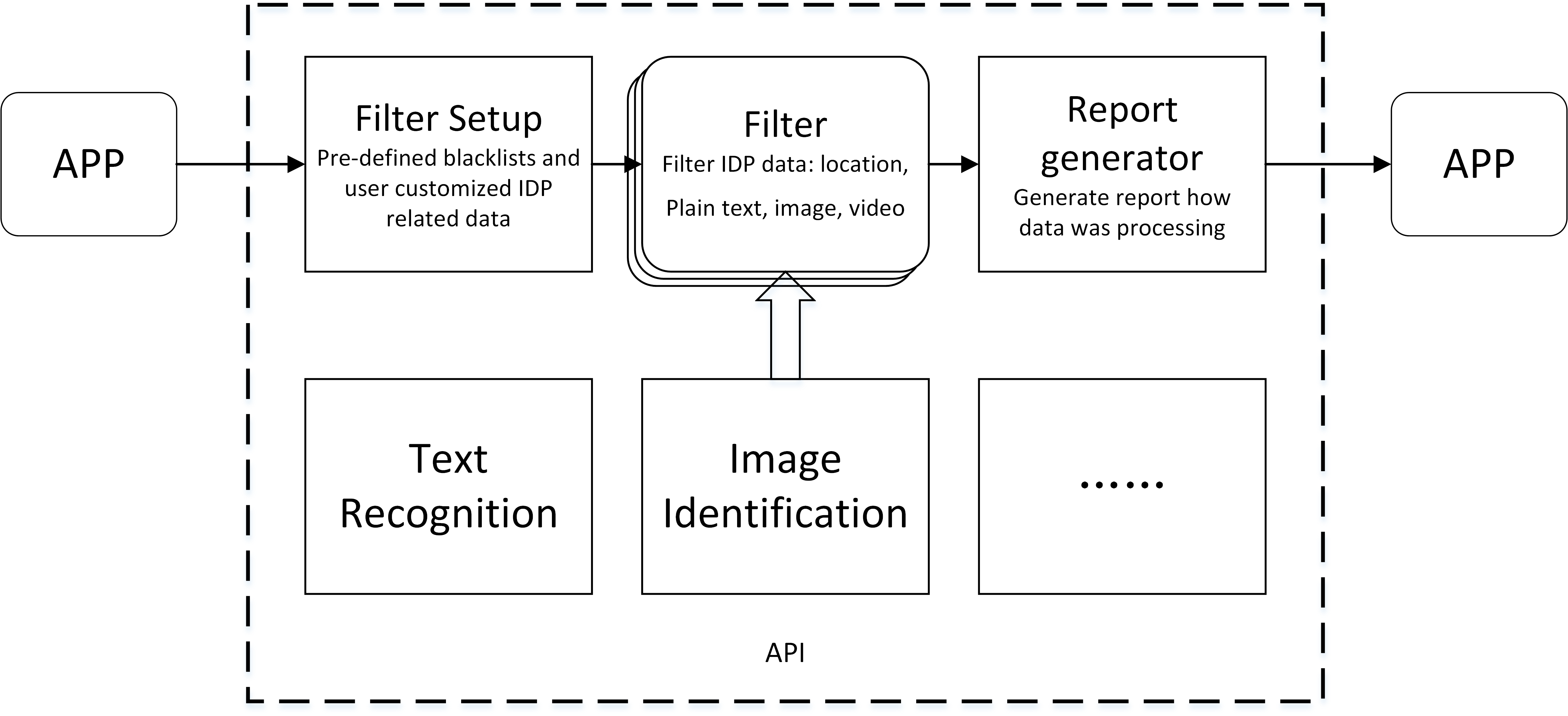}
\vspace{-3mm}
\caption{Overview of the IDPFilter API framework} 
\label{fig:API structure}
\vspace{1mm}
\end{figure}

We propose IDPFilter, an API designed to mitigate the risks associated with IDP in third-party applications. Drawing inspiration from established filtering methods, such as those evident in our proof-of-concept Flask-driven application (see Section \ref{sec:api}), this API endeavors to provide a robust layer of protection (see Figure~\ref{fig:User-APP-API2.0}. IDPFilter dynamically filters and masks sensitive information based on user preferences and predefined categories, thereby restricting inadvertent interdependent data leaks (see Figure~\ref{fig:API structure}). By marrying user-defined customization with system-driven presets, our solution encapsulates a forward-thinking approach to IDP. The crux of our system lies in its ability to improve on automated interdependent privacy safeguards, ensuring more efficient, customized protection for interdependent data and a trustworthy environment for app users.

The system consists of 4 main entities: users, applications, Interdependent Privacy Filters (IDPF), and Interdependent Privacy Settings Collector (ISC). IDPF will detect the content uploaded by the user, and if the data is IDP-related and matches the filter settings made by the user, IDPF will filter it. ISC collects IDP filter settings from the user. The default setting is to opt-in to the filter service. Of course, users have different needs in different scenarios, and ISC gives fine-grained choices, which means users can define how the filter actually works. IDPF and ISC together form IDPFilter. %We will introduce the specific implementation of this solution to the IDP problem in detail later.   
%We will give a complete implementation in the following chapter.

%\begin{figure}[tb]
%\centering
%\includegraphics[width=0.9\textwidth]{figures/API3rd_6.0.png}
%\vspace{-3mm}
%\caption{When sending information outside the trust boundary, IDP information will be filtered} 
%\label{fig:API3rd_6.0}
%\vspace{1mm}
%\end{figure}

\subsection{IDPFilter: mitigation scheme}
Our IDP mitigation scheme comprises three distinct interactions with regard to a given pair (app, user) during IDPFilter's operation (see Fig.~\ref{fig:finalComplete data flow}):

\begin{enumerate}

\item \textbf{Filter setup. }Initially, the IDP service provides a standard API to all interested third-party applications, encompassing various filtering schemes under different constraints. Third-party application developers can select and integrate the services that best align with their requirements. The Filter Setup can be done at any time the users use the service.

\item \textbf{Granting permissions and customization. }When users opt to install a third-party application, the app has to request a set of permissions related to IDPFilter. These permissions include i) whether the user allows the IDPFilter service to filter the information they share with the app in the future and ii) the extent of information filtering, e.g., \emph{whether the user permits others to share her information}; if not or only partially, the user can establish her own blacklist (whitelist), ensuring that the information referred to in the list is (not) filtered when other IDPFilter users share her information via the app in question. Note that the user only has to compile a blacklist (whitelist) once, and it is valid for all data transmission by the single app in question flowing through IDPFilter. User-compiled lists are further elaborated in Section~\ref{sec:poc_struct}.

\item \textbf{Data Flow. }Upon completing the first two stages, users share information through the third-party application, which utilizes the IDPFilter service. Based on the granted permissions and the chosen filtering scheme, the information is filtered before being shared with other users or third parties\footnote{third parties are stakeholders outside of users, app provider, and IDPFilter} as in Figure~\ref{fig:User-APP-API2.0}.
\end{enumerate}
\section{Proof-of-Concept: IDPTextFilter API}
\label{sec:api}

Users share multi-modal data via third-party applications, including text, voice, images, and videos. Technologies for image/voice/video recognition exist that could enable us to flag and reduce IDP-related leaks in all information modalities above. Nevertheless, in this section, we focus on basic textual data to illustrate the mechanism of mitigating IDP issues.

The reasons why we chose to focus on text filtering over other forms of data are the following:
\begin{enumerate}
    \item \textbf{Ubiquity and relevance. }Textual data is the most common type of data people share and interact with online. From social media posts to emails to articles, text is everywhere. Hence, text is indeed a relevant data modality for our proof-of-concept (PoC).
    \item \textbf{Simplicity and speed. } Text processing is generally simpler and faster than processing images or video. It requires less computational power and storage, which has enabled us to focus on demonstrating IDPFilter in action and avoid making processing cost and efficiency our primary concern.
    \item \textbf{Maturity of technology. } Natural Language Processing (NLP) has matured significantly over the past decades. We have advanced tools and libraries at our disposal that can be used directly to process text based on natural languages effectively and without developing algorithms from scratch. This has allowed for a streamlined development of our PoC.
\end{enumerate}

IDPFilter can be deployed as a web service and hosted by us or any trusted providers. Following the design principles P1-P5 (see Section~\ref{sec:design_princ}), we provide a series of interfaces available to \emph{any} third-party application; this ensures that IDP protection is available and hassle-free so that app developers can use it easily as opposed to coding their own IDP protection functionality. The app needs to
call the API to realize the function of IDP text filtering. If app developers integrate the IDPFilter API into their app, then users can both i) customize and ii) enjoy IDPFilter's features. From the users' perspective, APIs are presented as a (graphical) user interface integrated into the app. Therefore, users do not have to have technical knowledge in order to be protected by IDPFilter.

\subsection{IDPTextFilter: structure}
\label{sec:poc_struct}

Our PoC is comprised of the User Interface, the Privacy Settings Collector, the Text Filter, and three types of user-defined lists. Note that the User Interface we developed serves only demonstration purposes; it is not strictly a part of the IDP(Text)Filter. In a real-world deployment, the User Interface part is implemented \emph{within} the app.

\noindent \textbf{User Interface (UI). }The user interface provides an intuitive and seamless gateway for individuals to navigate and adjust their interdependent privacy settings (Principle P4: usability). Tailored to be user-friendly, this front-end component ensures that users have direct access and control over their personal data preferences.

\noindent \textbf{Text Filter (TF). }The Text Filter is responsible for dynamically applying the privacy settings and matching filter expressions, decided by the user, onto data that is being accessed by an app. By referencing parameters from the Privacy Settings Collector, TA ascertains which information can be shared and which should be withheld. TF is also responsible for generating filtering reports that form the basis of user feedback.

\noindent \textbf{Interdependent Privacy Settings Collector (ISC). }The Privacy Setting Collector plays the crucial role of gathering, storing, and updating the interdependent privacy settings of the users. To implement the core functionality of allowing users to choose which data to filter, the ISC design hinges on three foundational lists: the Self-Regarding Blacklist, the Other-Regarding Blacklist, and the Self-Regarding Whitelist. When integrated into the PoC, these lists ensure configurability and proper user incentives, resonating with Principles P2 and P3.

\noindent \textbf{Self-Regarding Blacklist (SRB). }A dynamic and customizable list where a user (User 1) designates private attributes that the app handles, but she does not want them to be shared via the app by any other user. As a data flow \emph{from another user using any app integrating IDPFilter containing information on User 1} passes through TA, this list acts as a primary checkpoint in flagging and removing any matches from getting to the app, other users, and third parties. There is a single SRB for every (user, app) pair. Note that entries of the SRB are i) limited to data handled by the app in question and ii) verified by the app. For example, in the case of the user's e-mail address, the app requests confirmation from the user and then verifies whether the user entered the same e-mail address to the SRB.

\noindent \textbf{Other-Regarding Blacklist (ORB). }Recognizing the intertwined nature of digital privacy, the ORB is dedicated to safeguarding the privacy of other individuals related to the user creating the list (e.g., friends, family, or simply other users). The ORB enables users who are willing to protect others' privacy to achieve this effectively. When a user enters a sensitive word that might be related to others' privacy into its ORB, the text sent by this user (only this user) using the respective app will be filtered accordingly. There is a single ORB for every (user, app) pair.

\noindent \textbf{Self-Regarding Whitelist (SRW). }Serving as a counterpoint to the SRB, this is where users earmark data segments they deem shareable \emph{by other users of the same app}. The SRW ensures the smoothness of information transmission when users agree to others sharing their information. There is a single SRW for every (user, app) pair.

By allowing users to define their own blacklists and whitelists, our prototype provides granular control over data sharing. As the user inputs data, the API checks it against these lists to filter out any sensitive or non-consented information. This reduces the potential risks associated with interdependent privacy in third-party applications.

\subsection{IDPTextFilter: implementation}
We have developed the PoC using Python\footnote{Code available at \url{https://github.com/shuai20/IDP_Filter}}. We utilized the Flask framework\footnote{\url{https://flask.palletsprojects.com/en/3.0.x/}} for development, accompanied by an HTML interface for displaying the privacy dashboard. We utilized the FlashText algorithm, which performs well when keywords are complete and exhibits linear complexity in relation to the length of the search text. This makes it particularly useful for many keywords, as all can be matched simultaneously in a single pass over the input string~\cite{DBLP:journals/corr/abs-1711-00046}. Each user can register to select different text filtering schemes and save their preferences. The TA itself offers a sensitive vocabulary database, categorized into names, links, countries, disease names, and street names. Users may opt for more stringent filtering schemes, such as filtering all numerals. Additionally, users can establish their own sensitive word blacklist, SRB and ORB. 

\begin{figure}[p]
\centering
\includegraphics[width=0.8\textwidth]{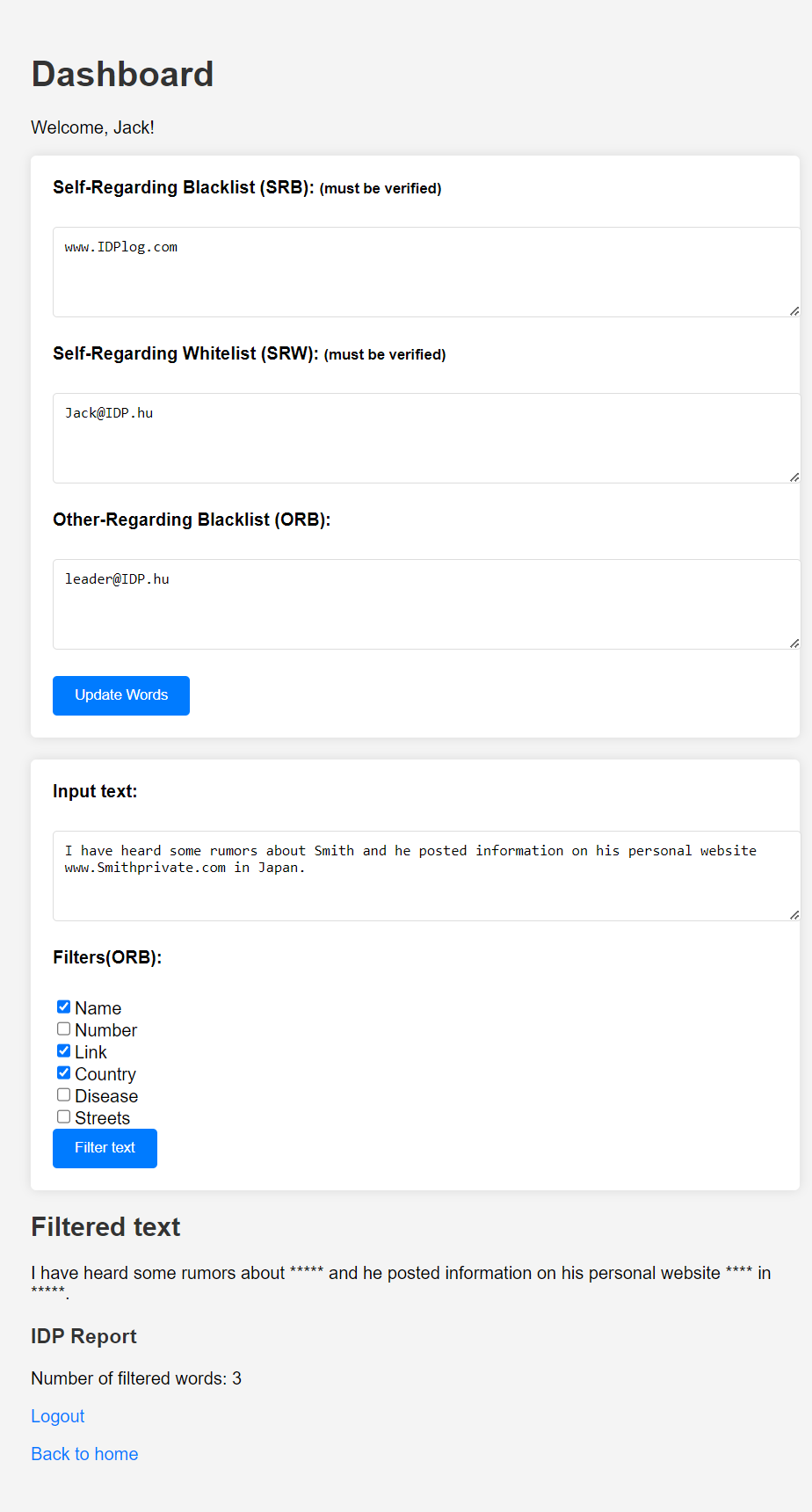}
\vspace{-3mm}
\caption{PoC: User Interface (dashboard)} 
\label{fig:PocDashboardscreenshot}
\vspace{1mm}
\end{figure}

Vocabulary in the SRB is added to a public blacklist, effective for all registered users, ensuring terms from the SRB are filtered during IDPTextFilter usage. On the other hand, ORB vocabulary is only effective for the individual user: only this user will have her own ORB words filtered when using the service. In addition, users can create a sensitive word whitelist SRW and incorporate it into their filtering scheme. SRW words are given the highest priority, ensuring their preservation regardless of the selected filtering scheme.
The UI is a user-centric dashboard designed for both i) user input of interdependent privacy settings and configurations and ii) feedback on filtering to the user. Users are greeted with personalized messages at a glance, ensuring a tailored experience. Users can effortlessly update their lists of sensitive and non-sensitive words via an intuitive mechanism. Beyond personal data settings, the dashboard provides robust filtering options, allowing users to specify categories of data they wish to shield, such as names, numbers, links, etc. After processing, the dashboard showcases a clear representation of the filtered text and provides a concise report on the number of words filtered. Navigation is streamlined with easily accessible options for logging out or returning to the homepage. Overall, the dashboard seamlessly bridges user preferences with advanced privacy features, ensuring fine-grained data protection, see Figure~\ref{fig:PocDashboardscreenshot}.

\subsection{IDPTextFilter: User operation}
\begin{figure}[p]
\centering
\includegraphics[width=0.9\textwidth]{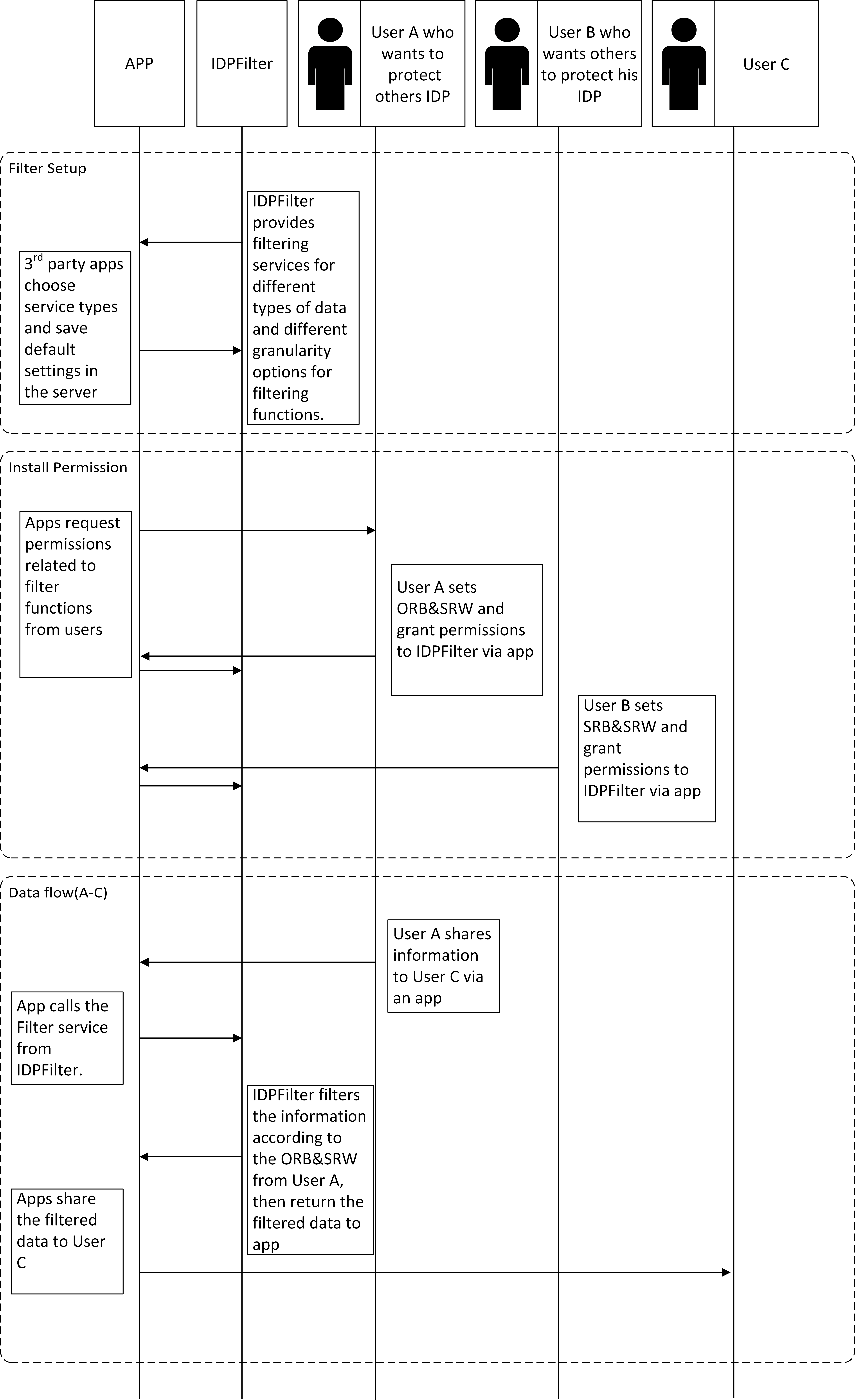}
\vspace{-3mm}
\caption{PoC data flow} 
\label{fig:finalComplete data flow}
\vspace{1mm}
\end{figure}

The operation of this prototype is mainly composed of three mechanisms: registration, text filter setup, and text filtering, as seen in Figure~\ref{fig:finalComplete data flow}.

\subsubsection{Registration. }
The prototype handles user registration and login based on a username and password pair. The password is not stored directly; the PoC rather transforms it using the Password-Based Key Derivation Function (PBKDF2) before storing it locally\footnote{using the \texttt{pbkdf2hmac()} function of the \texttt{hashlib} package in python}, implementing the current best practice in password storage. During user login, the user submits her username and password through the login form, and then the prototype calculates its PBKDF2 transform and matches it to the stored value in its SQLite database.

\subsubsection{Filter setup and execution. }
%\Url{https://github.com/shuai20/IDPAPI _23331}
We consider three users in this system: Alice, Bob, and Jack. Alice wants to share a piece of text information with Bob through an app. The text contains content related to Jack's personal information. Alice, Bob, and Jack can all add content to the SRB. In the content Alice sends to Bob, the content related to Jack's SRB will be filtered out. Alice can add sensitive words in the ORB that may involve Jack's personal information to protect Jack's privacy proactively. Jack can add content to the SRW, which is equivalent to announcing that others are allowed to share certain information involving Jack himself. SRB, SRW, and ORB constitute each user's text processing preferences. Users can modify them at any time. IDPFilter saves the user's current settings to ensure that users do not need to reconfigure each time.

%\afterpage{\clearpage}

\subsection{IDPTextFilter: Initial Evaluation}

We briefly evaluate the PoC from three aspects: privacy, filtering accuracy, and filtering efficiency.

\noindent \textbf{Privacy. }While it is not possible to quantify the IDP risk reduction of IDPTextFilter at this stage, we believe it is important to summarize its privacy-enhancing characteristics and provide a brief \emph{qualitative} evaluation. 

IDPTextFilter provides both transparency and the ability to control how one's private textual data is shared by others (i.e., interdependent privacy) in the context of third-party applications. In terms of risk, the increased transparency of data sharing/filtering (delivered in the form of detailed reports via the UI) enables users to estimate IDP risks more accurately. On the other hand, providing a fine-grained control mechanism (composed of blacklists/whitelists and a text-filtering component) enables users to effectively reduce both the likelihood and impact aspects of IDP risks. It is important to emphasize that IDPFilter caters both to individual users and the whole user base of an app by operationalizing both own and other-regarding privacy preferences.

\begin{figure}[tb]
\centering
\includegraphics[width=0.8\textwidth]{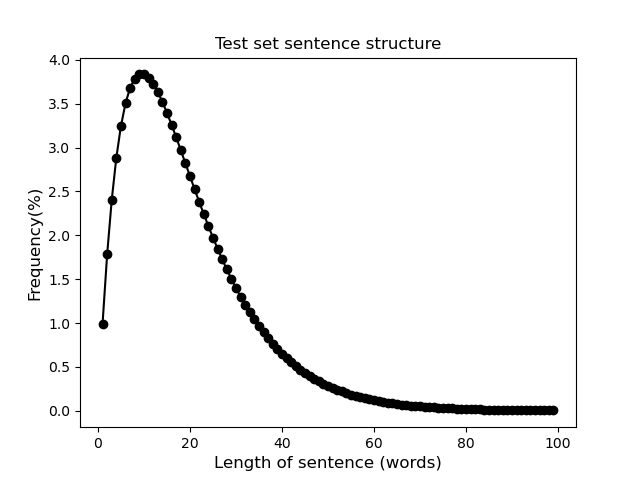}
\vspace{-3mm}
\caption{Empirical PDF of
sentence length (words) in each test set} 
\label{fig:SentenceStructure}
\vspace{-5mm}
\end{figure}

\noindent \textbf{Accuracy. } 
Although implementing the perfect pattern-matching tool is out of the scope of this paper, to illustrate the feasibility of the PoC, we devised accuracy tests for IDPTextFilter. To be able to perform an initial performance evaluation, we set up a local test environment (AMD Ryzen 7 4800H CPU 2.90 GHz, 32GB RAM with Windows 10 Education 22H2). Compared to traditional pattern-matching tools, the main feature to test is the users having different requirements for IDP protection; even for an individual user, requirements might change every time he/she shares information. The solution should be able to deal with multiple requirements and respond to modification quickly. For instance, every time users share data with others, they might choose to filter different categories like surnames or URLs, customize the blacklists and whitelists, or both. To check whether the accuracy is satisfactory, we implemented three different filter methods: regular expression, KMP algorithm~\cite{DBLP:journals/corr/abs-2204-08331}, and FlashText~\cite{2017arXiv171100046S}. For further comparison of filter results, we also implemented another Named Entity Recognition (NER) script by using the spacy library in Python~\footnote{\url{https://pypi.org/project/spacy/}}. Spacy NER has been proven to be an accurate and efficient tool to recognize the named entities in the text; in the following, we show the comparison of filtering results made by IDPTextFilter and Spacy NER. Spacy NER now could recognize limited types of entities such as PERSON, ORG, or MONEY; therefore, it cannot fully implement the filtering requirements of IDPFilter (e.g., a user would like to filter three street names in Budapest), but it can provide an industry-grade baseline for comparing the results when the same sentences are processed with the same IDP-related words. To construct the test sets,  we picked $10 \times 10^5$ random sentences from the Brown corpus\footnote{\url{http://korpus.uib.no/icame/brown/bcm.html}}. To make these sentences realistic, in each set, sentence lengths conform to the Zipf distribution ($ f_{\text{exp}} = 1.1 \cdot L^{1} \cdot 0.90^{L}$)~\cite{sigurd2004word} visualized in Figure \ref{fig:SentenceStructure}.

\begin{table}[tb]
\centering
\caption{Number of words filtered in 10 sentence sets:  IDPFilter (800 most common surnames) vs. NER (PERSON)}
\label{tab: filterd words number}
\begin{tabular}{p{3.1cm}p{0.8cm}p{0.8cm}p{0.8cm}p{0.8cm}p{0.8cm}p{0.8cm}p{0.8cm}p{0.8cm}p{0.8cm}p{0.8cm}}
\hline
          & S1   & S2   & S3   & S4   & S5   & S6   & S7   & S8   & S9   & S10  \\ \hline
IDPFilter(Regex) & 1455 & 1394 & 1432 & 1424 & 1390 & 1241 & 1394 & 1360 & 1304 & 1364 \\ \hline
IDPFilter(KMP) & 1455 & 1394 & 1432 & 1424 & 1390 & 1241 & 1394 & 1360 & 1304 & 1364 \\ \hline
IDPFilter(FlashText) & 1455 & 1394 & 1432 & 1424 & 1390 & 1241 & 1394 & 1360 & 1304 & 1364 \\ \hline
Spacy NER & 2255 & 2097 & 2272 & 2258 & 2227 & 2068 & 2161 & 2195 & 2044 & 2223 \\ \hline
\end{tabular}
\end{table}

Due to the inflexibility of NER, we chose to recognize (filter) PERSON entities and compare the result when IDPTextFilter filters the name category. We show the number of filtered words for the four different methods in Table~\ref{tab: filterd words number} with the $800$ most common English surnames as the IDP-related blacklist.

After checking the filtered sentences, we can draw two conclusions. First, IDPTextFilter accurately filters out all the words in the list. On the contrary, NER makes mistakes when recognizing people's names. Some names are not recognized, and some other entities are mistakenly recognized as names.
Second, NER is trained on a large amount of data; therefore, it performs well in approximate string matching and can identify fictitious names and nicknames, while IDPTextFilter cannot identify names outside the data set.
In summary, IDPTextFilter can properly filter text based on user preferences if fine-grained blacklists are provided. Nevertheless, future work might consider adding NER-inspired approximate matching capabilities for (semi-)automatically creating sensible filtering rules; note that approximate expressions might create IDP issues themselves.

\begin{figure}[tb]
\centering
\includegraphics[width=0.9\textwidth]{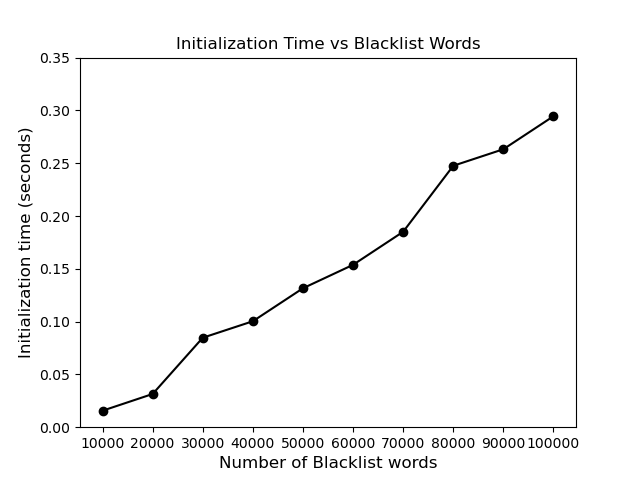}
\vspace{-3mm}
\caption{Initialization time cost} 
\label{fig:Init time}
\vspace{-5mm}
\end{figure}

\begin{figure}[tb]
\centering
\includegraphics[width=0.9\textwidth]{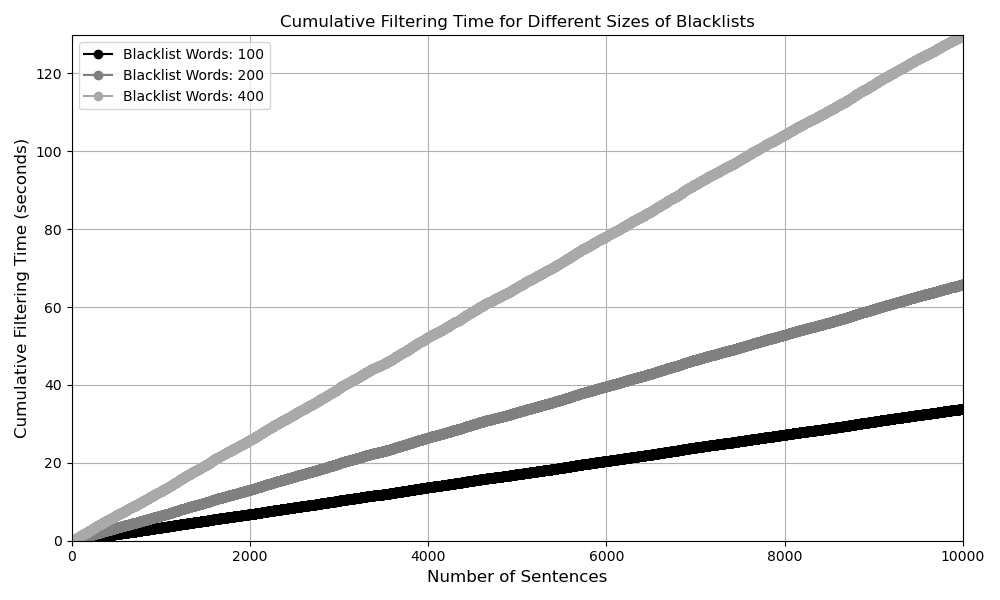}
\vspace{-3mm}
\caption{Time cost (KMP) with different number of blacklist entries} \label{fig:time cost kmp}
\vspace{-5mm}
\end{figure}

\begin{figure}[tb]
\centering
\includegraphics[width=0.9\textwidth]{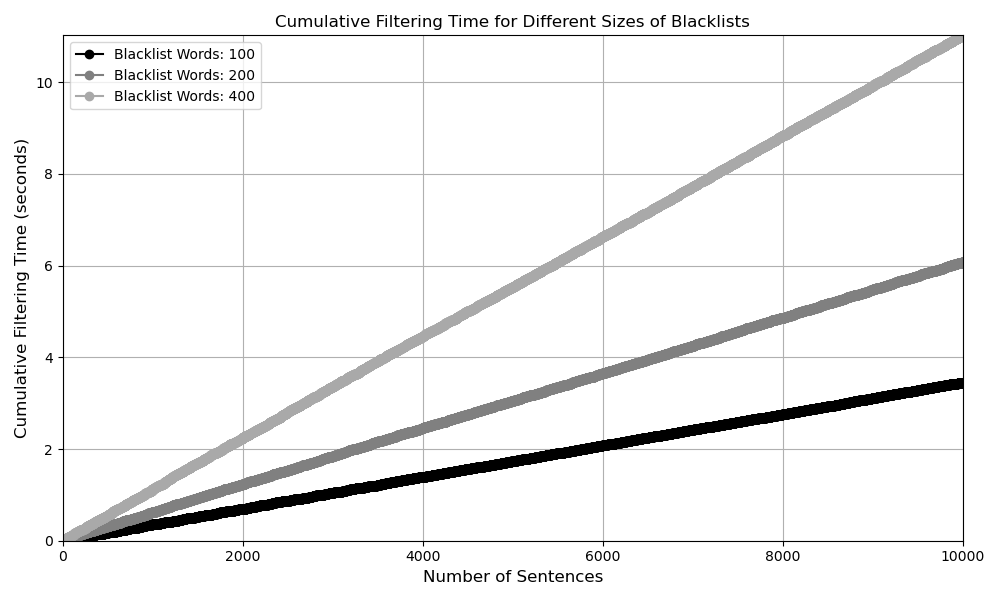}
\vspace{-3mm}
\caption{Time cost (FlashText) with different number of blacklist entries} \label{fig:time cost Flashtext}
\vspace{-5mm}
\end{figure}

%\newpage
\noindent \textbf{Efficiency. }
Considering that information sharing occurs all the time, mitigating the IDP issue requires an efficient solution that can promptly process a large amount of information.  We first utilize the Knuth-Morris-Pratt (KMP) algorithm~\cite{DBLP:journals/corr/abs-2204-08331} to improve efficiency. To quantify efficiency, first, we show that when IDPFilter is provided with a large dataset to filter, the initialization time of the system is sufficiently short. We generated multiple blacklists ($10,000$ to $100,000$ words per set) for measuring the initialization time. We obtained the ratio of the time required for pre-processing to the size of the blacklist text; see Figure~\ref{fig:Init time}.
Then, we obtain the time costs when IDPTextFilter processes the test sets with IDP-related blacklists of sizes $100$, $200$, and $400$. Figure~\ref{fig:time cost kmp} shows the average filtering time costs for $10,000$-sentence test sets: $33.27 s$, $67.28 s$, and $127.34 s$, respectively. The FlashText solution we ultimately adopted is far superior to other solutions in terms of efficiency. The time cost of filtering sentences under the same conditions is only about one-tenth of that of other solutions. Figure~\ref{fig:time cost Flashtext} shows the average filtering time costs for $10,000$-sentence test sets: $3.45 s$, $6.22 s$, and $11.95 s$, respectively. Note that IDPTextFilter is designed to reload the user's filter settings every time it is invoked for the user, so the filtering time values shown also include $10,000$ initializations.
Note that we did not take further steps to improve the efficiency of IDPTextFilter as the prototype's performance was already sufficient to justify its promise regarding real-world deployment. Naturally, turning IDPTextFilter into a carrier-grade service would require more performance optimization.

\section{Conclusion}
\label{sec:conclusion}

\noindent\textbf{Summary. }In this paper, we carried out a comprehensive study on the interdependent privacy issues that inherently plague third-party applications in general. Specifically, our three research questions regarding RQ1) the pervasiveness of the privacy interdependence issues across platforms, RQ2) the collateral information collection of real-life apps, and RQ3) the possibility of designing and implementing a common mitigation mechanism were answered affirmatively. First, we provided data and reasoning about the proliferation of privacy interdependence on app platforms (RQ1) and in existing apps (RQ2). Then, most importantly, we designed IDPFilter, a platform-agnostic API that enables privacy-conscious app providers \emph{and} end-users to filter out collateral information, i.e., data collected from certain app users but (also) implicating other natural persons as data subjects. We also provided a proof-of-concept implementation, IDPTextFilter, that realizes the proposed filtering logic on textual data (RQ3). 

\noindent\textbf{Benefits of IDPFilter. }IDPFilter was constructed with five design principles in mind. First, it provides rich IDP-related information to the end-user to improve transparency and raise awareness. Second, the filtering mechanism is configurable to balance between privacy and app utility carefully. Third, IDPFilter can be used voluntarily, yet it aligns with the economic
incentives of both app providers and end-users. Fourth, IDPFilter is platform-agnostic and easy to use for all stakeholders, providing a usable one-stop shop for IDP mitigation. Last, realizing that the complete re-design of third-party app permission systems is out of our reach, IDPFilter is compatible with the current mechanisms of app platforms, not requiring any major change regarding how apps and platforms operate.

\noindent\textbf{Limitations. }Naturally, our IDP mitigation mechanism has some limitations. First, from a trust perspective, the entity providing the IDPFilter service observes all collected data. This can be improved in different ways: i) using advanced cryptographic solutions, such as homomorphic/functional/searchable encryption that enables computation on encrypted data, or ii) by letting the app platforms themselves deploy the filtering service. Second, we provide neither a full-fledged implementation (realizing filtering on other, more complex data modalities) nor a comprehensive performance evaluation of our proof-of-concept prototype, IDPTextFilter. We believe that our contributions are sufficient to answer RQ3 affirmatively and leave the comprehensive treatment regarding both a full implementation and performance evaluation to future work. Third, recently brought into the foreground by the increasing popularity of applied large language models, we acknowledge that simple text filtering cannot remove all potential privacy implications owing to the correlated nature of natural languages: sensitive information can be revealed even if complete lines of text were removed~\cite{brown2022does}. All in all, IDPFilter cannot mitigate all IDP issues emerging in the third-party app context; nevertheless, it provides a first-of-its-kind solution that is able to improve the current situation considerably\cite{DBLP:conf/esorics/LiuHB21}.

\section*{Acknowledgments}
Project no. 138903 has been implemented with the support provided by the Ministry of Innovation and Technology from the NRDI Fund, financed under the FK\_21 funding scheme.

%\begin{figure}[tb]
%\centering
%\includegraphics[width=0.9\textwidth]{figures/ZoomCat.PNG}
%\vspace{-3mm}
%\caption{Zoom Marketplace permission system structure} 
%\label{fig:Zoomcat}
%\vspace{1mm}
%\end{figure}

\color{black}
%Cores of the mitigation: 1. Let users forbid others from sharing their own private data. 2. When users know they are going to share others' private data and would like to protect others' privacy, help them filter the data. 3. When users want to protect others' privacy but can not be sure whether the data is related to others' privacy, help them identify and filter the data. 

%\newpage
\bibliographystyle{unsrt}
\bibliography{reference}
\end{document}